\documentclass[twocolumn,showpacs,superscriptaddress,twoside,prx]{revtex4-1}
\usepackage{amsthm}
\usepackage{amsmath}
\usepackage{amssymb}
\usepackage{graphicx}
\usepackage{url}
\usepackage{dsfont}
\usepackage{float}
\usepackage{bm}
\usepackage{ifthen}
\usepackage[usenames,dvipsnames]{color}
\usepackage{mathrsfs}
\usepackage[colorlinks=true,citecolor=blue,urlcolor=black]{hyperref}
\usepackage{float}
\usepackage{afterpage}
\usepackage{subfigure}
\usepackage{adjustbox}
\usepackage{times}
\usepackage{wrapfig}
\usepackage{multirow}
\newcommand{\sket}[1]{{\ensuremath{\lvert#1\rangle}}}
\newcommand{\lket}[1]{{\ensuremath{\left\lvert#1\right\rangle}}}
\newcommand{\ket}[1]{\if@display\lket{#1}\else\sket{#1}\fi}
\newcommand{\sbra}[1]{{\ensuremath{\langle#1\rvert}}}
\newcommand{\lbra}[1]{{\ensuremath{\left\langle#1\right\rvert}}}
\newcommand{\bra}[1]{\if@display\lbra{#1}\else\sbra{#1}\fi}
\newcommand{\sbraket}[2]{{\ensuremath{\langle#1\rvert#2\rangle}}}
\newcommand{\lbraket}[2]{{\ensuremath{\left\langle#1\!\left\rvert\vphantom{#1}#2\right.\!\right\rangle}}}
\newcommand{\braket}[2]{\if@display\lbraket{#1}{#2}\else\sbraket{#1}{#2}\fi}

\newcommand{\sketbra}[2]{{\ensuremath{\lvert #1\rangle\!\langle #2\rvert}}}
\newcommand{\lketbra}[2]{{\ensuremath{\left\lvert #1\right\rangle\!\!\left\langle #2\right\rvert}}}
\newcommand{\ketbra}[2]{\if@display\lketbra{#1}{#2}\else\sketbra{#1}{#2}\fi}

\theoremstyle{plain}

\theoremstyle{definition}

\usepackage{color}
\definecolor{myred}{rgb}{1,0,0}

\definecolor{myblue}{rgb}{0,0,0.8}

\definecolor{myyellow}{rgb}{0.9,0.8,0}

\definecolor{mygreen}{rgb}{0,0.6,0}

\definecolor{myorange}{rgb}{0.6,0.6,0}

\definecolor{mycerul}{rgb}{0,0.6,1}

\newcommand{\be}{\begin{equation}}
\newcommand{\ee}{\end{equation}}

\newcommand{\ignore}[1]{} 

\begin{document}
\title{  Hardware-Efficient Bosonic  Quantum Error-Correcting  Codes Based on Symmetry Operators}

\author{Murphy Yuezhen Niu}
\affiliation{Research Laboratory of Electronics, Massachusetts Institute of Technology, Cambridge, Massachusetts 02139, USA}
\affiliation{Department of Physics, Massachusetts Institute of Technology, Cambridge, Massachusetts 02139, USA}
\author{Isaac L. Chuang}
\affiliation{Research Laboratory of Electronics, Massachusetts Institute of Technology, Cambridge, Massachusetts 02139, USA}
\affiliation{Department of Physics, Massachusetts Institute of Technology, Cambridge, Massachusetts 02139, USA}
\affiliation{Department of Electrical Engineering and Computer Science, Massachusetts Institute of Technology, Cambridge, Massachusetts 02139, USA}
\author{Jeffrey H. Shapiro}
\affiliation{Research Laboratory of Electronics, Massachusetts Institute of Technology,  Cambridge, Massachusetts 02139, USA}
\affiliation{Department of Electrical Engineering and Computer Science, Massachusetts Institute of Technology, Cambridge, Massachusetts 02139, USA}
\date{\today}

\begin{abstract}

We establish a symmetry-operator framework for designing  quantum error correcting~(QEC) codes based on fundamental properties of the underlying system dynamics. Based on this framework, we  propose three hardware-efficient bosonic QEC codes  that are suitable for $\chi^{(2)}$-interaction based quantum computation in multi-mode-Fock-bases:  the  $\chi^{(2)}$ parity-check code, the $\chi^{(2)}$  embedded error-correcting code, and  the $\chi^{(2)}$ binomial code.   All of these QEC codes detect photon-loss or photon-gain errors by means of photon-number parity measurements, and then correct them via $\chi^{(2)}$ Hamiltonian evolutions and linear-optics transformations.  Our symmetry-operator framework provides a systematic procedure for finding QEC codes that are not stabilizer codes, and it enables convenient extension of a given encoding to higher-dimensional qudit bases.  The $\chi^{(2)}$ binomial code is of special interest because, with $m\le N$ identified from channel monitoring, it can correct $m$-photon loss errors, $m$-photon gain errors, and $(m-1)$th-order dephasing errors using logical qudits that are encoded in $O(N)$ photons.  In comparison, other bosonic QEC codes require  $O(N^2)$ photons to correct the same degree of bosonic errors. Such improved photon-efficiency underscores the additional error-correction power that can be provided by channel monitoring. We develop quantum Hamming bounds for photon-loss errors in the code subspaces associated with the $\chi^{(2)}$ parity-check code and the $\chi^{(2)}$ embedded error-correcting code, and we prove that these codes saturate their respective bounds.  Our $\chi^{(2)}$ QEC codes exhibit hardware efficiency in that they address the principal error mechanisms and exploit the available physical interactions of the underlying hardware, thus reducing the physical resources required for implementing their encoding, decoding, and error-correction operations, and their universal encoded-basis gate sets.
 
\end{abstract}

\maketitle

\section{Introduction}
Quantum error-correcting~(QEC) codes are essential for realizing large-scale quantum computation. QEC codes protect quantum information by encoding each logical computational-basis state into a higher-dimensional physical subspace in a manner that permits errors to be detected and corrected.  The first QEC codes were generic codes, i.e., they made no assumptions about the underlying hardware~\cite{Shor1995,Perfectcode1996,Bennett1996,Steane,GottesmanThesis}, hence their error models do not exploit hardware-specific biases towards particular errors.  The achievable code rates of generic QEC codes are therefore constrained by fundamental limits, such as the quantum Hamming bound and the quantum singleton bound~\cite{GottesmanThesis}.   Consequently, a generic QEC code demands more encoding overhead than would be necessary when the quantum computation is run on hardware with a small set of dominant errors, rather than the full error set assumed by that code.   This overhead excess impedes implementation of large-scale quantum computation as compared to what could be accomplished with a \emph{hardware-efficient} QEC code, viz., one that is matched to the   chosen physical implementation.
  
Compared to generic QEC codes, hardware-efficient QEC codes offer a quicker route to the break-even point~\cite{LiangNature2016}, at which encoded quantum information is retained beyond the coherence time of its physical constituents. To do so they exploit the available physical interactions to correct system-specific errors in a low-encoding-overhead manner that does not lavish resources on correction capability for unlikely errors~\cite{Leghtas2013,LiangNature2016,Gambetta2017}.  
Hardware-efficient codes that protect the same amount of quantum information as generic codes thus use fewer controlled-Hamiltonian evolutions, measurements,  and classical controls in their encoding, decoding, and error correcting of quantum information, and in their universal gate-set constructions. They also introduce fewer error mechanisms and offer higher code rates than generic QEC codes.

Hardware-efficient QEC codes are especially relevant for bosonic quantum computation, in which photons are the information carriers. Examples include quantum optical computation using Kerr nonlinearities~\cite{Chuang1995}, linear-optical quantum computation~\cite{KLM}, and continuous-variable quantum computation~\cite{Seth,Mirrahimi2014}, among others. Because photons are prone to loss, and photon-photon interactions are extremely weak, bosonic QEC codes focus on correcting photon-loss errors using very limited forms of photon-photon interactions while striving to be hardware efficient.  The bosonic codes from Refs.~\cite{Chuang1995,Chuang1997} correct up to $N$-photon-loss errors by using linear optics and Kerr nonlinearities~(four-wave-mixing) to encode each logical qubit into two bosonic modes with up to $N^2$ photons in each mode.   The quantum parity-check codes~ \cite{Gilchrist2005,Ralph2007,Ludkin2014} use $N^2$ single photons distributed over $2N^2$ modes to correct $N$-photon-loss errors with measurement-induced universal gates.  The GKP codes~\cite{Gottesman2001} use Kerr nonlinearities and atom-photon coupling to encode logical qudits into superpositions of squeezed states that approximately correct for lowest-order photon-loss errors, which are regarded as that architecture's likely errors.  

The development of cat codes~\cite{Munro1999, Ralph2004,Leghtas2013,Kirchmair2013,Vlastakis2013,Mirrahimi2014,Sun2014,LiangNature2016} represents an important step toward hardware efficiency.  
Cat codes are bosonic QEC codes tailored to the promising quantum-computing architecture whose physical qubits are microwave photons stored in superconducting resonators.  Cat codes' universal gate set relies on induced four-wave mixing interactions in Josephson junctions, which are much stronger than optical four-wave mixing in Kerr media.  Cat codes have lower encoding overhead than generic QEC codes, because they introduce fewer error mechanisms.  In addition, their requiring fewer physical resources than generic QEC codes  makes it easier for them to reach the break-even point~\cite{LiangNature2016}.
 
The cat codes' success invites the following question:  can we design hardware-efficient bosonic QEC codes for the quantum computation scheme based on three-wave-mixing?  Langford \emph{et al}.~\cite{Zeilinger2011} were the first to point out the possibility of realizing universal quantum computation in the single-photon qubit basis using $\chi^{(2)}$ interactions for coherent photon conversion together with linear-optics transformations.  Quantum computation using only these resources is of interest because the $\chi^{(2)}$ interaction is a lower-order nonlinearity, hence potentially stronger than four-wave mixing.  Furthermore, exciting new technologies---such as solid-state circuits ~\cite{Long2008}, flux-driven Josephson-junction parametric amplifiers~\cite{Devoret2010,Bergeal2010,Oliver2016}, superconducting resonator arrays~\cite{Tian2011,Tian2012}, ring resonators~\cite{Sipe2007}, and frequency-degenerate double-lambda systems~\cite{TwoLambda}---have been expanding the platforms for and increasing the efficiencies of $\chi^{(2)}$ interactions. Hardware-efficient QEC codes for Langford's protocol~\cite{Zeilinger2011}  would establish the feasibility of this emerging quantum computing scheme.  Such codes would also add to existing toolkits for some four-wave-mixing approaches to quantum computation, because $\chi^{(2)}$ interactions can be realized by means of four-wave mixing with a strong, nondepleting pump~\cite{Zeilinger2011,Jennewein2014,meyer2015power}.

Unfortunately, the single-photon qubit basis used in Ref.~\cite{Zeilinger2011} is not closed under $\chi^{(2)}$ Hamiltonian evolutions, nor is it suitable  for error correction, because any single-photon-loss error will destroy the quantum information carried by the qubit.  Moreover, existing  single-photon, multi-mode,  QEC codes---such as the bosonic code~\cite{Chuang1995}, the quantum parity-check code~\cite{Ralph2004}, and the NOON code~\cite{Loock2016}---are not designed for hardware-efficient operation on an underlying architecture comprised of $\chi^{(2)}$ interactions and linear optics.

We took a first step toward overcoming the preceding difficulties in Ref.~\cite{Niu2017} by showing that universal quantum computation using only $\chi^{(2)}$ interactions and linear optics could be realized with a multi-mode Fock basis in an irreducible subspace of the $\chi^{(2)}$ Hamiltonian, i.e., the closed subspace $\mathcal{H}_N$ of quantized single-mode signal ($s$), idler ($i$), and pump ($p$) states spanned by multi-mode Fock states whose photon numbers $\{n_k: k = s,i,p\}$ satisfy $(n_s+n_i)/2 + n_p = N$.  In the present paper we take the next step, by proposing the first QEC codes for $\chi^{(2)}$-based universal quantum computation and then demonstrating their hardware efficiency.  To do so we establish a symmetry-operator framework, that leverages the symmetry of the  physical subspace supporting the logical codewords and the symmetry of the measurable syndromes. 

We  choose a multi-mode Fock basis spanning an irreducible subspace of $\chi^{(2)}$ Hamiltonian evolutions~\cite{Niu2017} as our physical qudit basis. This choice ensures closed dynamics during quantum computation, and hence avoids leakage errors in the absence of photon loss or gain.  The measurable syndromes include the photon-number parity measurements and generalized photon-number parity measurements.   For each code, symmetry operators shared by both the physical code subspace and the measurable syndromes   are carefully chosen to stabilize the logical basis states.   Error operators that do not commute with the logical basis' symmetry operators can thus be detected through appropriate syndrome measurements. Our use of symmetry operators affords a systematic procedure to find  QEC codes that are not conventional stabilizer codes, and also enables the extension of our $\chi^{(2)}$ QEC codes to arbitrary qudit dimensionalities and different numbers of bosonic modes.

We expect photon loss or gain to be the dominant error mechanisms for our $\chi^{(2)}$ architecture, so our three hardware-efficient codes---the $\chi^{(2)}$ parity-check code~($\chi^{(2)}$ PCC), the $\chi^{(2)}$ embedded error-correcting code~($\chi^{(2)}$ EECC), and the $\chi^{(2)}$ binomial code~($\chi^{(2)}$ BC)---are tuned for such errors.   The $\chi^{(2)}$ PCC is so named because its second physical qudit provides a parity check on the first qudit;   the EECC is so named because we embed an $ N $-dimensional logical qudit into a  $2N-1$ dimensional physical qudit; and  the $\chi^{(2)}$  BC is so named because it uses conjugated binomial symmetry in its construction. Each code has its own merit in regards to hardware efficiency.  The $\chi^{(2)}$ PCC  has a constant code rate for logical qudits of any dimension.  It  corrects single-photon-loss and single-photon-gain errors, and it detects dephasing errors. The $\chi^{(2)}$ EECC corrects single-photon-loss and single-photon-gain errors and has the highest code rate of our three codes.  The $\chi^{(2)}$ BC is our most powerful code, when sufficient resources are available.  Using $O(N)$ photons for its encoding, it corrects $m$-photon loss errors, $m$-photon gain errors---but not mixtures of loss and gain errors---and dephasing errors up to the $(m-1)$th order, given an $m\le N$ value identified from channel monitoring that identifies the error order but not its type~\cite{Salles2008}.  As a result, despite the $\chi^{(2)}$ PCC and the $\chi^{(2)}$ EECC's capabilities being limited to single-photon errors, they are more promising than the $\chi^{(2)}$ BC for near-term experimental demonstration because they have explicit universal gate-set implementations and error-correction procedures, and they do not require channel monitoring.    

It is worth emphasizing that the $\chi^{(2)}$ BC is the first Fock-basis bosonic QEC code that can correct $N$-photon-loss errors using $O(N)$ photons for its encoding; all previous bosonic QEC codes with that error-correction capability require encoding with $O(N^2)$ photons~\cite{Chuang1995,Gottesman2001,Gilchrist2005,Ludkin2014,Loock2016,Girvin2016,Munro1999,Ralph2004,liang2016cat}.  Being able to correct the loss of a constant fraction of the total photons for any code size is of great advantage for both large-scale quantum computation and  long-range quantum communication. However, the $\chi^{(2)}$ BC requires channel-monitoring resources---which our other codes do not---that it uses to determine the number of photons that have been lost or gained. Our result thus highlights the importance of channel monitoring for the extra error-correction power it can provide by obviating a constraint from the error-correction condition~\cite{knill1997theory}, and it alerts us to the need for resource-efficient channel monitoring.

Assuming the universal gates realized through $\chi^{(2)}$ interactions have no errors in themselves, correctable photon-loss/gain errors do not induce additional logical errors~(physical qudit rotation errors) in our multi-mode Fock-basis encoding. Moreover, the encoding, decoding, error correction and universal gates are all realizable with just $\chi^{(2)}$ interactions and linear optics. 

To establish the optimalities of the $\chi^{(2)}$ PCC and the $\chi^{(2)}$ EECC with respect to their code rates, we  develop generalized quantum Hamming bounds for any $[[n\log_2(q), k\log_2 (b) , 2t+1]] $ code, i.e., one that encodes $k$ logical qudits of dimension $b$ into $n$ physical qudits of dimension $q$ and corrects either $t$-physical-qudit rotation errors or $t$-photon-loss errors.  Then we show that the  $\chi^{(2)}$ PCC and the  $\chi^{(2)}$ EECC saturate their respective bounds for photon-loss errors.  In doing so, we find that the  quantum Hamming bounds for photon-loss errors give much higher code rates than those for physical-qudit rotation errors.  This disparity arises from our use of three-mode Fock states, for which certain photon-loss errors move the original code subspace to a higher-dimensional subspace.  That increased subspace dimension facilitates a more efficient error correction procedure, leading to qubit-basis code rates of 1/2 for the $\chi^{(2)}$ PCC and $1/\log_2(3)$ for the $\chi^{(2)}$ EECC, as opposed to the 1/5 code rate of the generic qubit-basis QEC code for qubit rotation errors.  Furthermore, our multi-mode encodings do not require all bosonic modes to have the same loss rate, something that is necessary for other multi-mode bosonic codes.

We begin in Sec.~\ref{HardwareEfficiency} by presenting a performance comparison between our three hardware-specific bosonic codes, and the bosonic (but otherwise generic) GKP codes, establishing a quantitative rationale for hardware efficiency.  We develop in Sec.~\ref{encodingsection} a symmetry-operator framework---inspired by the familiar stabilizer codes---and use it to define  the $\chi^{(2)}$ PCC, EECC, and BC codes in the context of the coherent-$\chi^{(2)}$ physical model.   Because these codes reside in subspaces of the physical model's full Hilbert space, it is important for quantum computation to still have universality within the encoded bases, and we show how this can be done in Sec.~\ref{sec:univencod}.  Finally, for perspective on our codes' performance relative to ultimate limits, we present quantum Hamming bounds, generalized to accommodate our codes' qudit bases, in Sec.~\ref{hamming}, before concluding in Sec.~\ref{Conclusion}.


\section{Hardware Efficiency}~\label{HardwareEfficiency}

A hardware-efficient QEC code~\cite{Leghtas2013,LiangNature2016} for a given physical architecture should minimize the physical resources employed for its encoding, decoding, and error-correction operations, and for its universal gate sets in the encoded basis.  At the same time, it should avoid introducing additional error mechanisms, and it should have a high code rate.  The physical resources of concern in this regard include controlled Hamiltonian evolutions, measurements, and classical controls.  Thus we shall quantify a QEC code's hardware efficiency in terms of six metrics:  the physical resources it requires for encoding, decoding, error correction, and universal computation; its dominant error mechanisms; and its code rate. Below, after some additional information about our three $\chi^{(2)}$  QEC codes, we compare their hardware-efficiency metrics against those of the widely studied GKP code.
 
The hardware efficiency of $\chi^{(2)}$ QEC codes is facilitated by their code subspaces being irreducible subspaces of $\chi^{(2)}$ Hamiltonian evolutions, viz., they are closed under such evolutions. This choice conserves photon-number parity, and enables universal gate sets to be realized with just $\chi^{(2)}$ interactions and linear optics.  It follows that all $\chi^{(2)}$ QEC codes require just generalized photon-number parity measurements~\cite{Sun2014,Mirrahimi2014} for their error-detection operations.  
  
\begin{table} 
\centering
\begin{tabular}{c @{\hspace{0.2cm}}  c @{\hspace{0.2cm}} c @{\hspace{0.2cm}} c @{\hspace{0.3cm}} c@{\hspace{0.3cm}} c@{\hspace{0.3cm}} c} \hline \hline
  & $n$ &   $q$ & $b$ & correctable & total photons\\\hline\hline
 $\chi^{(2)}$ PCC & $2$   & $N$ & $N$ &\shortstack{single-photon loss\\single-photon gain} &$ 3(N-1) $\\\hline
 $\chi^{(2)}$ EECC   &   1 & $(2N - 1)  $ & $ N $ &\shortstack{single-photon loss\\single-photon gain}   & $  3(N-1)  $\\
\hline
 $\chi^{(2)}$ BC  & $1$ &$ 2N $ & 2& \shortstack{$m$-photon loss\\$m$-photon gain\\  dephasing\footnote{$(m-1)$th order dephasing errors.   $ m\leq N $.} }&$ 3(N-1/2) $ \\
\hline \hline
\end{tabular} 
\caption{Comparison of the correctable error sets and total number of photons required for our $\chi^{(2)}$ QEC codes, all of which encode 1 logical qudit of dimension $b$ into $n$ physical qudits of dimension $q$.} \label{Table1}
\end{table}
 
Table.~\ref{Table1} compares code dimensions, correctable error sets, and average total  photon numbers of the $\chi^{(2)}$ PCC, the $\chi^{(2)}$ EECC, and the $\chi^{(2)}$ BC.  For all three codes, the required total number of  photons scales linearly with the physical basis dimension $q$ because they use  constant photon-number spacing within their three-mode Fock-state bases. 
 
We define the code rate of an $[[n\log_2(q), k\log_2(b),2t+1]]$ code to be $ k\log_2(b)/n\log_2(q) $, which reduces to the familiar $k/n$ code rate for qubit encoding ($b=2$) into physical qubits ($q=2$)~\cite{GottesmanThesis}.  Thus, because the $\chi^{(2)}$ PCC is a $[[2\log_2(N), \log_2(N),3]]$ code, its code rate is 1/2 and it corrects single-photon (loss or gain) errors. The $\chi^{(2)}$ EECC is a $[[\log_2(2N-1), \log_2(N),3]]$ code, so it too corrects single-photon (loss or gain) errors, but its code rate, $\log_2(N)/\log_2(2N-1)$, is higher than that of the $\chi^{(2)}$ PCC and approaches unity as $N$ grows without bound.  The $\chi^{(2)}$ BC has the lowest code rate, $1/\log_2(2N)$, of our three codes.  With channel monitoring that identifies $m\le N$, it can correct $m$-photon (loss or gain) errors as well as  $(m-1)$th-order dephasing errors.
\begin{table} 
\centering  
\begin{tabular}{|c|c|c|c|l}
\cline{1-4}
&  \multicolumn{1}{ |c| }{GKP code} & \multicolumn{1}{ |c| }{\shortstack{\\ $\chi^{(2)} $ PC\\ $\chi^{(2)}$ EECC} } &\multicolumn{1}{ |c| }{ $\chi^{(2)}$ BC} \\ \cline{1-4}
\multicolumn{1}{ |c| }{ \shortstack{errors\\mechanisms}} &  \shortstack{\\
GEE   \\ photon loss \\  dephasing errors }
    & \shortstack{  
  photon loss\\  dephasing errors  } & \shortstack{  
  photon loss\\  dephasing errors  } &    \\ \cline{1-4}     
\multicolumn{1}{ |c| }{ \shortstack{ encoding\\resources}} & \shortstack{\\ $\chi^{(3)}$  \\linear optics\\$ \chi^{(2)} $\footnote{The GKP code's $\chi^{(2)}$ resources are all incoherent $\chi^{(2)}$ interactions (see text). }    }  &\shortstack{linear optics\\$\chi^{(2)}$  }&\shortstack{linear optics\\$\chi^{(2)}$  }&  \\ \cline{1-4}
\multicolumn{1}{ |c| }{ \shortstack{ decoding\\resources}} & \shortstack{\\ homodyne \\ linear optics\\$\chi^{(2)}$   }   &\shortstack{linear optics\\$\chi^{(2)}$\\PNR}& PNR &   \\ \cline{1-4}
\multicolumn{1}{ |c| }{ \shortstack{\shortstack{error-\\ correction\\resources}}} &\shortstack{\\ ancilla state\footnote{Ancilla state is the equal superposition of the logical computational-basis states, $(\ket{\tilde{0}}+\ket{\tilde{1}})/\sqrt{2}$, whose generation requires an additional GKP-encoding resource.}\\ homodyne\footnote{Only realizes approximate error correction. }\\$\chi^{(2)}$ }  & \shortstack{linear optics\\$\chi^{(2)}$ \\GPNP } &\shortstack{\\CM\\linear optics\\$ \chi^{(2)} $ \\GPNP }&     \\ \cline{1-4}
\multicolumn{1}{ |c| }{ \shortstack{ encoded\\ universality\\resources}} & \shortstack{\\ PNP\\  $\chi^{(2)}$   \\ feedforward control} & \shortstack{linear optics\\$\chi^{(2)}$  }  &\shortstack{linear optics\\$\chi^{(2)}$  }&     \\ \cline{1-4}
\multicolumn{1}{ |c| }{code rate} &\shortstack{ max[$g((1-\gamma)N)$ \\ $-g(\gamma N), 0]$}\footnote{GKP-code channel capacity for photon-loss errors under a mean photon-number $N$ constraint~\cite{Wilde2016}, where $\gamma$ is the photon-loss probability and $g(x)\equiv (1+x)\log_2(1+x)-x\log_2(x) $.  }& \shortstack{\\ $1/2$\footnote{$\chi^{(2)}$ PCC's code rate.} \\$ 1/\log_2(3)  $\footnote{$\chi^{(2)}$ EECC's code rate.}}    &$1/\log_2(2N)$&      \\ \cline{1-4}
\end{tabular}
\caption{Hardware-efficiency metrics for the GKP code~\cite{Gottesman2001}, the $\chi^{(2)}$  PCC, the $\chi^{(2)}$ EECC, and the $\chi^{(2)}$ BC when all four encode a single logical qubit.  GEE:  Gaussian embedded error.   PNR:  photon-number-resolving detection.  PNP:   photon-number parity measurement.   GPNP:  generalized photon-number parity measurement~\cite{Sun2014}. CM: channel monitoring.}
\label{Table2}
\end{table}

Table.~\ref{Table2} compares our hardware-efficiency metrics for the GKP code~\cite{Gottesman2001}, the $\chi^{(2)}$  PCC, the $\chi^{(2)}$ EECC, and the $\chi^{(2)}$ BC when all four encode a single logical qubit.  The salient points of this comparison are as follows.

\noindent \textit{Error Mechanisms}

Each logical-basis state in the GKP code~\cite{Gottesman2001} is encoded into a superposition of squeezed states.  Ideally, these states should have infinite squeezing, but a practical GKP-code realization must have finite energy, and hence finite squeezing. Consequently, its logical-bases states are not  orthogonal, causing what are known as Gaussian embedded errors (GEEs).  GEEs, in turn, induce logical errors through the GKP code's use of  SUM gates.   GKP codes are also susceptible to photon-loss and dephasing errors. 

For our $\chi^{(2)}$ QEC codes, on the other hand, the dominant error mechanisms are photon loss and dephasing errors. Low-order photon-loss errors do not cause additional logical errors,  which greatly eases the resource burden on error correction for $\chi^{(2)}$ QEC codes.  

\vspace{10pt}

\noindent\textit{Resources for Encoding and Decoding}

The GKP code's encoding operation requires  $\chi^{(2)}$ and $\chi^{(3)}$ interactions plus linear optics.  The $\chi^{(2)}$ interactions, which are used to generate squeezing, are all \emph{incoherent}, i.e., they have a strong, nondepleting pump (treated as a classical resource) and weak (quantum-mechanical) signal and idler modes.  

Our $\chi^{(2)}$ QEC codes' encoding requires only $\chi^{(2)}$ interactions and linear optics.  Here, however, the principal $\chi^{(2)}$ interactions needed are \emph{coherent}, i.e., the signal, idler, and pump modes are all quantum mechanical. (See Sec.~\ref{coherentinteract} for more information about coherent $\chi^{(2)}$ interactions.)

The decoding resources required by the four codes in Table~\ref{Table2} are comparable.

\vspace{10pt}

 \noindent\textit{Resources for Error Correction}

The GKP code realizes \textit{approximate} error correction using ancillae states prepared in the equal superposition of the logical-basis states,  homodyne measurements, and incoherent $\chi^{(2)}$ interactions.  Ancillae preparation requires $\chi^{(3)}$ interactions that add to the GKP code's error-correction resource burden.
 
In comparison, the $\chi^{(2)}$  PCC and $\chi^{(2)} $ EECC  perform \textit{exact} error correction using coherent and incoherent $\chi^{(2)}$ interactions,  linear optics and generalized photon-number parity measurements.  We provide the error correction circuits for qutrit-basis $\chi^{(2)} $ PCC  and the qubit basis  $ \chi^{(2)}$ EECC using just these resources in Appendices~\ref{ENCODE} and~\ref{GATE}, respectively.     

The $\chi^{(2)}$ BC, on the other hand, requires additional channel monitoring, which makes it only applicable to architectures---like superconducting resonator arrays---in which such monitoring the total number of photon lost to the environment is possible.  Thus it will not be hardware efficient when channel monitoring poses a major implementation burden.

\vspace{10pt}

\noindent\textit{Resources for Encoded Universality}
   
The  GKP code requires photon-number parity measurements, incoherent $\chi^{(2)}$ interactions, and feedforward controls to implement a universal gate set in its encoded basis. As shown in~\cite{Sanders2007}, conventional $\chi^{(2)}$ crystals cannot be pumped hard enough to get GKP-code gates of fidelity sufficient to exceed the error-correction threshold.

The universal encoded-basis gate sets for our $\chi^{(2)}$ QEC codes employ coherent $\chi^{(2)}$ interactions and linear optics, thus they too are currently precluded by the limited nonlinearity of conventional $\chi^{(2)}$ crystals.   However, as noted earlier, new technologies are emerging~\cite{Long2008,Devoret2010,Bergeal2010,Oliver2016,Tian2011,Tian2012,Sipe2007,TwoLambda}  that may afford the strong nonlinearity required for coherent $\chi^{(2)}$ interactions, and these may also enable the strong squeezing that the GKP code needs for its universal encoded-basis gate set.

\vspace{10pt}

\noindent\textit{Code Rate}

Because the GKP code uses a continuous-variable physical basis for its encoding, we are using the  quantum channel capacity, max[$g((1-\gamma)N) -g(\gamma N), 0]$,  under a mean photon-number $ N $ constraint~\cite{Wilde2016} with photon-loss probability $\gamma$ and $g(x)\equiv(1+x)\log_2(1+x)-x\log_2(x) $, as an upper bound on its code rate. As yet, however, whether the GKP code is capacity achieving has not been determined. The code rates of the $\chi^{(2)}$ PCC and the $\chi^{(2)}$ EECC that encode a single logical qubit are constant, whereas the code rate of the $\chi^{(2)}$ BC decreases as its error-correction capability increases.  A fair comparison between these codes' rates has yet to be obtained, as it requires analysis of channel fidelity under the same error model~\cite{Victor2017}.

\vspace{10pt}

\noindent\textit{Hardware-Efficiency Summary}

Overall, as compared to the GKP code, our $\chi^{(2)} $ QEC codes reduce the amount of physical resources necessary for the encoding, decoding, error correction, and universal gate implementations without introducing new error mechanisms. These resource reductions are natural consequences of our having customized the $\chi^{(2)}$  QEC codes to the underlying  $\chi^{(2)}$-interaction computational hardware.


\section{$\chi^{(2)}$ Quantum Error-Correcting Codes}
\label{encodingsection}

We define and present in detail our three hardware-efficient bosonic quantum error-correction codes in this section, beginning in Sec.~\ref{subsec:chitwo} with a brief elaboration of the hardware primitives and the coherent $\chi^{(2)}$ interactions employed.  Our codes are not traditional stabilizer QEC codes; instead, they rely on a different---but stabilizer-inspired---method of symmetry operators, which takes advantage of natural symmetries available in the physical model.  The formalism for this symmetry-operator method is detailed in Sec.~\ref{subsec:symmetryop}.  These foundations then enable us to present, in increasing order of complexity and capability, the $\chi^{(2)}$ parity-check code (in Sec.~\ref{chi2PCC}), the $\chi^{(2)}$ embedded error-correcting code (in Sec.~\ref{chi2EECC}), and the $\chi^{(2)}$ binomial code (in Sec.~\ref{subsec:chi2bc}).

The presentation of each code focuses on the code's symmetry operators, defining the code basis states, explaining what errors are corrected, and giving the mathematical and physical rationale for why the errors are detectable and correctable.  Details giving physical procedures, employing coherent $\chi^{(2)}$ interactions, for encoding, decoding, detecting errors, and correcting errors, are deferred to Appendices~\ref{ENCODE} and \ref{GATE}.

\subsection{Coherent $\chi^{(2)}$ Interactions\label{coherentinteract}} 
\label{subsec:chitwo}

Before embarking on QEC code designs for $\chi^{(2)}$ quantum computation, it behooves us to elaborate on the hardware primitives with which their encoding, decoding, and error-correction operations, and their universal logical-basis gate sets are implemented.   These consist of coherent and incoherent $\chi^{(2)}$ interactions and linear-optics transformations.  The necessary linear-optics transformations are phase shifters, dichroic mirrors, ordinary and polarizing beam splitters.  The incoherent $\chi^{(2)}$ interactions we need are nondepleting-pump, frequency-degenerate, type-II phase-matched spontaneous parametric downconversion (SPDC), and  quantum-state frequency conversion (QFC)~\cite{Kumar1990,Albota2004,Albota2006}. Coherent $\chi^{(2)}$ interactions are the core, however, of our architectures~\cite{Niu2017} for multi-mode Fock basis quantum computation and hence also for our $\chi^{(2)}$ QEC codes. The particular coherent $\chi^{(2)}$ interactions we will require are: frequency-degenerate, type-0 and type-I phase-matched SPDC, in which a single-photon Fock state at pump frequency $\omega_p=2\omega$ is converted into a two-photon Fock state at frequency $\omega$; second-harmonic generation (SHG), in which a two-photon Fock state at frequency $\omega$ is converted to a single-photon Fock state at frequency $2\omega$; and sum-frequency generation (SFG), in which orthogonally-polarized, single-photon Fock states at frequency $\omega$ are converted into a single-photon Fock state at frequency $2\omega$.  More importantly, for this paper's purposes, the inherent symmetry properties of coherent $\chi^{(2)}$ interactions provide easy routes for embedding a lower-dimensional logical basis into a higher-dimensional physical basis as discussed later in this section.

Specifically, the $\chi^{(2)}$ Hamiltonians for coherent interactions between single-mode signal, idler, and pump fields are linear combinations of  $\hat{G}_1$ and  $\hat{G}_2$ terms given by
\begin{align}\label{HSPDC11}
	\hat{G}_1
		&=\frac{i\kappa}{2}\left[\hat{a}_s^{\dagger}\hat{a}_i^{\dagger}\hat{a}_p - \hat{a}_s\hat{a}_i\hat{a}_p^{\dagger}\right],\\
		\hat{G}_2
		&=\frac{\kappa}{2}\left[\hat{a}_s^{\dagger}\hat{a}_i^{\dagger}\hat{a}_p + \hat{a}_s\hat{a}_i\hat{a}_p^{\dagger}\right],
\label{HSPDC}
\end{align}
with $\{\hat{a}_k^{\dagger}: k = s,i,p\}$ being the photon-creation operators of the signal, idler, and pump, and the real-valued $\kappa$ being the interaction strength.  For our purposes, the signal and idler will always be taken to have frequency $\omega$ and the pump will always be assumed to have frequency $2\omega$, but their polarizations will depend on what computational primitive is being implemented, i.e., we may use SPDC with type-0 (co-polarized signal, idler, and pump) or type-I (orthogonally-polarized signal and idler, pump co-polarized with the idler) phase matching.  The full Hilbert space of three-mode states satisfies, $\mathcal{H} = \oplus_{N=0}^\infty \mathcal{H}_N$, where the direct sum is over the Hamiltonian's $N$-pump-photon irreducible subspaces, $\mathcal{H}_N \equiv {\rm Span}\{\ket{0,0,N},\ket{1,1,N-1},\ldots,\ket{N,N,0}\}$, with $\ket{n_s, n_i, n_p}$ denoting a Fock state having $n_s$ photons in the signal mode, $n_i$ photons in the idler mode, and $n_p$ photons in the pump mode~\cite{Niu2017}.

\subsection{Symmetry Operators}
\label{subsec:symmetryop}

Crucial to our $\chi^{(2)}$ QEC codes' ability to detect and correct photon-loss and photon-gain errors is the symmetry properties of each $N$-pump-photon subspace, $\mathcal{H}_N$.  Specifically,   every state  $\ket{\psi}=\sum_{n=0}^N c_n\ket{n,n,N-n}$ in  $\mathcal{H}_N$ obeys the following eigenvalue-eigenstate relations:
\begin{align}\label{symmetry1}
&(\hat{n}_s+\hat{n}_p)\ket{\psi}=N\ket{\psi},\\\label{symmetry2}
&(\hat{n}_i+\hat{n}_p)\ket{\psi}=N\ket{\psi},\\\label{symmetry3}
&(\hat{n}_s -\hat{n}_i)\ket{\psi}=0,
\end{align}
where $\hat{n}_k \equiv \hat{a}_k^\dagger\hat{a}_k$, for $k = s,i,p$.
Consequently, the photon-number parity vector, 
\begin{align}\label{pvectdefn}
{\bf p} & \equiv [\langle \hat{n}_s + \hat{n}_i\rangle, \langle \hat{n}_s+\hat{n}_p \rangle,  \langle \hat{n}_i+\hat{n}_p\rangle]{\rm mod}\,2,\\
 &= [2N, N, N]{\rm mod}\,2,
\end{align}
is constant for any $|\psi\rangle \in \mathcal{H}_N$. Such symmetry  derives from energy conservation within $\mathcal{H}_N$. 

To establish a deeper understanding of the preceding symmetry properties, we introduce symmetry operators for $\mathcal{H}_N$.  The three-mode Fock-state basis, $\{\ket{n,n,N-n} : 0\le n \le N\}$, for $\mathcal{H}_N$ is characterized by these basis states'  invariance under the application of  symmetry operators $\hat{Z}_{s,p}^{(N+1)}=e^{i2\pi/(N+1)}\hat{Z}_s^{(N+1)}\otimes \hat{Z}_p^{(N+1)} $ and $\hat{Z}_{i,p}^{(N+1)}= e^{i2\pi/(N+1)}\hat{Z}_i^{(N+1)}\otimes \hat{Z}_p^{(N+1)} $. Here, \begin{align}
\hat{Z}_k^{(N+1)}\equiv \sum_{n=0}^Ne^{i2\pi n/(N+1)}\ket{n}_{k\,k}\bra{n},
\end{align}
where $\ket{n}_k$ for $n=0,1,\ldots,N$ is an $n$-photon Fock state of mode $k$, for $k = s,i,p$.  
If  $N+1$  is  a prime number,  $\hat{Z}_k^{(N+1)}$ is the mode-$k$ Pauli $Z$ operator for the qudit basis $\{\ket{0}_k,\ket{1}_k,\ldots,\ket{N}_k\}$.  We, however, do not require $N+1$ to be prime, because our $\chi^{(2)}$ QEC codes are not  stabilizer codes~\cite{Beth2003}. So, we refer to $\hat{Z}_{s,p}^{(N+1)}$ and $\hat{Z}_{i,p}^{(N+1)}$ as physical-subspace symmetry operators in $\mathcal{H}_N$.   Where unambiguous, we do employ some stabilizer terminology in our explanations, but we also make it clear below how our codes are distinct from traditional stabilizer codes.

In order to redundantly encode a lower-dimensional logical basis into a higher-dimensional physical basis, we need additional symmetry operators to stabilize the logical state:  within the code's physical subspace, only the simultaneous unity-eigenvalue eigenstates of all symmetry operators in the given set will be selected as logical-basis states.   

The $\chi^{(2)}$ PCC, which encodes an $N$-dimensional logical qudit into two $N$-dimensional physical qudits, first imposes the $\{\hat{Z}^{(N)}_{s_\ell,p_\ell},\hat{Z}^{(N)}_{i_\ell,p_\ell}: \ell =  1,2\}$ symmetries to restrict its code space to $\mathcal{H}_{N-1}^{\otimes 2}= \mathcal{H}_{N-1}^{(1)}\otimes \mathcal{H}_{N-1}^{(2)}$.  It then requires two additional symmetry operators for its construction.   The first additional symmetry operator is the photon-number inversion-symmetry operator $\hat{V}_1^{(N-1)}\otimes \hat{V}_2^{(N-1)}$, where 
\begin{align}\label{Voperator}
&\hat{V}_\ell^{(N-1)} \equiv \hat{V}_{s_\ell}^{(N-1)}\otimes \hat{V}_{i_\ell}^{(N-1)}\otimes\hat{V}_{p_\ell}^{(N-1)},
\end{align}
with
\begin{align}
& \hat{V}_{k_\ell}^{(N-1)} \equiv \sum_{n=0}^{N-1}\ket{N-1-n}_{k_\ell\,k_\ell}\bra{n},
 \end{align}
for $k=s,i,p$, inverting mode $k_\ell$'s qudit basis, viz.,  
\begin{align}
\hat{V}_\ell^{(N-1)}\ket{n,n,N-1-n}_\ell=\ket{N-1-n, N-1-n, n}_\ell.
\end{align}
The second additional symmetry operator we need for the $\chi^{(2)}$ PCC is the swap operator $\hat{X}^{(N)}_{1,2}$.  In $\mathcal{H}_{N-1}^{\otimes 2}$ it swaps three-mode basis states between the two subspaces, i.e., for $0\le n_k,n'_k \le N-1$ and $k = s,i,p$, 
\begin{align}
 \hat{X}_{1,2}^{(N)}\ket{n_s,n_i, n_p}_1\ket{n_s^\prime,n_i^\prime, n_p^\prime}_2=\ket{n_s^\prime,n_i^\prime, n_p^\prime}_1\ket{n_s,n_i, n_p}_2.
\end{align}

The $\chi^{(2)}$ EECC encodes a single logical qudit of dimension $N$ into a single physical qudit of dimension $2N-1$.  It only requires three symmetry operators for its encoding:  $\hat{Z}_{s,p}^{(2N-1)}, \hat{Z}_{i,p}^{(2N-1)}$ and $\hat{V}^{(2N-2)}$.  The $\chi^{(2)}$ BC encodes a single logical qubit into a single physical qudit of dimension $2N$.  It also requires only three symmetry operators for its encoding:  $\hat{Z}_{s,p}^{(2N)}, \hat{Z}_{i,p}^{(2N)}$ and $\hat{\Pi}_s\hat{U}_{\rm BS}\hat{V}^{(2N-1)}\hat{U}_{\rm BS}^\dagger$, where $\hat{U}_{\rm BS}$ is a pseudo-beam-splitter operator, to be described later, that operates on the physical-qudit subspace $\mathcal{H}_{2N-1}$, and 
\begin{equation}
\hat{\Pi}_s \equiv \sum_{n=0}^\infty (-1)^n|n\rangle_{s\,s}\langle n|
\end{equation}  
is the signal mode's parity operator.

Table~\ref{Table3} summarizes the symmetry operators used for our $\chi^{(2)}$ QEC codes. All three codes require the $\hat{Z}^{(M)}_{s,p}$ and $\hat{Z}^{(M)}_{i,p}$ operators, for appropriate $M$ values, to stabilize their logical states to their physical code-subspace:  $\mathcal{H}_{N-1}^{\otimes 2}$ for the $\chi^{(2)}$ PCC, $\mathcal{H}_{2N-2}$ for the $\chi^{(2)}$ EECC, and $\mathcal{H}_{2N-1}$ for the $\chi^{(2)}$ BC.   All three also require $\hat{V}^{(M)}$ operators, with appropriate $M$ values, for photon-number inversion symmetry.  However, because it uses two physical qudits for encoding, the $\chi^{(2)}$ PCC also requires the swap-symmetry operator, $\hat{X}^{(N)}_{1,2}$, between its two $N$-dimensional physical-basis subspaces.  

Note that $\hat{Z}^{(M)}_k$  is diagonal in the physical qudit basis $\{\ket{n,n,M-1-n}_k : 0\le n \le M-1\}$, so it does not commute with the photon-annihilation operators $\{\hat{a}_k\}$, or with the photon-creation operators $\{\hat{a}_k^\dagger\}$. Although $\hat{V}^{(M)}_k$ is not diagonal in the $\{\ket{n,n,M-n}_k : 0\le n \le M\}$ basis, it too fails to commute with the $\{\hat{a}_k\}$ and the $\{\hat{a}_k^\dagger\}$.  Likewise, the $\hat{X}_{1,2}^{(N)}$ operator, which is not diagonal in the $\{\ket{n_1,n_1,N-1-n_1}_1\ket{n_2,n_2,N-1-n_2}_2 : 0 \le n_1,n_2 \le N-1\}$ basis, also fails to commute with the photon-annihilation and photon-creation operators.  These commutation failures will lead to the $\chi^{(2)}$ PCC and $\chi^{(2)}$ EECC's being able to correct single-photon (loss or gain) errors, and the $\chi^{(2)}$ BC's being able to correct $m$-photon (loss or gain) errors and $(m-1)$th-order dephasing errors when $m \le N$ is identified by channel monitoring. 
  
\begin{table} [h]
\centering
\begin{tabular}{c|ccccc} \hline\hline
 $\chi^{(2)}$ PCC & $\hat{Z}_{s_\ell,p_\ell}^{(N)}$, & $\hat{Z}_{i_\ell,p_\ell}^{(N)}$,   & $\hat{V}_1^{(N-1)}\otimes \hat{V}_2^{(N-1)}$, & $\hat{X}^{(N)}_{1,2}$ &\\\hline
 $\chi^{(2)}$ EECC   &  $\hat{Z}_{s,p}^{(2N-1)}$, & $\hat{Z}_{i,p}^{(2N-1)}$, & $\hat{V}^{(2N-2)}$ &   &  \\
\hline
 $\chi^{(2)}$ BC  & $\hat{Z}_{s,p}^{(2N)}$, & $\hat{Z}_{i,p}^{(2N)}$, & $\hat{\Pi}_s\hat{U}_{\rm BS}\hat{V}^{(2N-1)}\hat{U}_{\rm BS}^\dagger$ &  &    \\
\hline \hline
\end{tabular} 
\caption{Symmetry operators used for constructing our three $\chi^{(2)}$ QEC codes.} \label{Table3}
\end{table}

Since the majority of our symmetry operators lie outside the qudit Pauli group, $\chi^{(2)}$ QEC codes  are not  stabilizer codes. Similar to what was shown in Ref.~\cite{Beth2003}, however, error operators that do not commute with all the symmetry operators nonetheless can be detected by nondemolition measurements of the symmetry operators with additional ancillae states.   Hence, the error-detection procedures for our $\chi^{(2)}$ QEC codes resemble those for stabilizer codes.  Nondemolition measurements of the photon-number parities
\begin{align}
 \hat{P}_{j,k}^{(N+1)}\equiv (-1)^{n_j+n_k }\ket{n_j}_{j\,j}\bra{n_j}\otimes \ket{n_k}_{k\,k}\bra{n_k},
\end{align} 
where $j\neq k$ are indices for different bosonic modes, have been realized with superconducting-resonator technology~\cite{Sun2014}. Equations~(\ref{symmetry1})--(\ref{symmetry3}) show that our $\chi^{(2)}$ QEC codes obey photon-number parity symmetry.  So, nondemolition measurements of the $\{\hat{P}_{j,k}^{(M)}\}$ for each code's appropriate $M$ value will provide a practical route to error detection.  

In general, symmetry properties are not obligatory for constructing QEC codes~\cite{Chuang1997}, but our symmetry-operator formalism  establishes a systematic framework for finding new QEC codes for available measurement schemes and physical subspace choices, and it offers a path for extending codes to arbitrary qudit bases.  Indeed, it is that pathway that motivates our exhibiting the use of symmetry properties to establish high-dimensional versions of the $\chi^{(2)}$ PCC and the $\chi^{(2)}$ EECC.  But,  because these codes can only correct single-photon loss or gain errors, their practical utility degrades as their dimensionality increases with fixed photon-loss and photon-gain probabilities, which motivates our development of the $\chi^{(2)}$ BC. Moreover, we note that the symmetry operators  used for constructing each code in Table.~\ref{Table3} commute with each other from the same set, and are either monomial matrices~\cite{Nest2011} or a monomial matrix conjugated by unitary transformations. Our findings thus call for a more general code-construction framework that generalizes the Pauli-stabilizer formalism to a larger class of error-correcting codes constructed by commuting monomial matrices and conjugated monomial matrices with analytically simple expressions for the quantum error correction conditions.

\subsection{$\chi^{(2)}$ Parity-Check Code}\label{chi2PCC}

For the $\chi^{(2)}$ PCC we draw inspiration from previous work~\cite{Chuang1997,aharonov2008fault, Ludkin2014},  and encode one $N$-dimensional logical qudit into two $N$-dimensional physical qudits for correcting single-photon loss or gain errors.   The $\chi^{(2)}$ PCC thus has code rate $1/2$ regardless of the encoded qudit's dimension, and it has minimum encoding overhead because it saturates the corresponding quantum Hamming bound, see Sec.~\ref{hamming}.  The resources needed for these functions are coherent and incoherent $\chi^{(2)}$ interactions, linear optics, and photon-number parity measurements, implying that there is a clear route to implementing the $\chi^{(2)}$ PCC given nonlinearity sufficient for the coherent $\chi^{(2)}$ interaction's second-harmonic generation (SHG), sum-frequency generation (SFG), and spontaneous parametric downconversion (SPDC) primitives.

To encode an $N$-dimensional logical qudit into two $N$-dimensional physical qudits, we first require that the logical qudits obey physical-subspace symmetry, i.e., they are unity-eigenvalue eigenstates of $\{\hat{Z}^{(N)}_{s_\ell,p_\ell},\hat{Z}^{(N)}_{i_\ell,p_\ell} : \ell =1,2\}$ and hence lie in $\mathcal{H}_{N-1}^{\otimes 2}$.  Next we impose photon-number inversion symmetry, by requiring the logical-basis states to also be unity-eigenvalue eigenstates of  $\otimes_{\ell = 1}^2\hat{V}_\ell^{(N-1)}$.  Finally, we require that the logical-basis states be invariant under the swap-symmetry operation, viz., under application of $\hat{X}^{(N)}_{1,2}$.  

As a simple example, we now show how the qutrit-basis $\chi^{(2)}$ PCC is obtained by the preceding procedure. In the qutrit $\chi^{(2)}$ PCC, each logical-qutrit basis state is encoded into two physical qutrits.  Imposing the physical-subspace symmetry characterized by $\{\hat{Z}^{(3)}_{s_\ell,p_\ell},\hat{Z}^{(3)}_{i_\ell,p_\ell}, : \ell = 1,2\}$ constrains the code to the subspace spanned by the nine unity-eigenvalue eigenstates of these operators, $\{\ket{n_1,n_1,2-n_1}_1\ket{n_2,n_2,2-n_2}_2 : 0 \le n_1,n_2 \le 2\}$, i.e., it constrains the code to $\mathcal{H}_2^{\otimes 2}$.  Enforcing  the photon-number inversion symmetry $\otimes_{\ell = 1}^2\hat{V}_\ell^{(N-1)}$ then reduces the symmetry subspace to the five-dimensional space spanned by 
\begin{align}
\{ &(\ket{0,0,2}_1\ket{0,0,2}_2+\ket{2,2,0}_1\ket{2,2,0}_2)/\sqrt{2}, \nonumber \\ 
&(\ket{2,2,0}_1\ket{0,0,2}_2+ \ket{0,0,2}_1\ket{2,2,0}_2)/\sqrt{2}, \nonumber \\  
& \ket{1,1,1}_1\ket{1,1,1}_2, \nonumber \\
& [(\ket{0,0,2}_1+\ket{2,2,0}_1)\ket{1,1,1}_2]/\sqrt{2}, \nonumber \\
& [\ket{1,1,1}_1(\ket{0,0,2}_2+ \ket{2,2,0}_2 )]/\sqrt{2}\}. \nonumber  
 \end{align}
Imposing the swap symmetry $\hat{X}^{(N)}_{1,2}$ reduces the symmetry subspace dimension to three, yielding the logical qutrit basis:
\begin{align}\label{EC1}
&\ket{\tilde{2} }=(\ket{2,2,0}_1\ket{0,0,2}_2 +\ket{0,0,2}_1\ket{2,2,0}_2)/\sqrt{2},\\
&\ket{\tilde{1} }=(\ket{2,2,0}_1\ket{2,2,0}_2 +\ket{0,0,2}_1\ket{0,0,2}_2)/\sqrt{2},\\\label{EC3}
&\ket{\tilde{0} }=\ket{1,1,1}_1\ket{1,1,1}_2.
\end{align}

It is straightforward to verify that the qubit-basis $\chi^{(2)}$ PCC's logical-basis states, 
\begin{align}\label{EC1qubit}
&\ket{\tilde{1} }=(\ket{1,1,0}_1\ket{0,0,1}_2 +\ket{0,0,1}_1\ket{1,1,0}_2)/\sqrt{2},\\\label{EC2qubit}
&\ket{\tilde{0} }=(\ket{1,1,0}_1\ket{1,1,0}_2 +\ket{0,0,1}_1\ket{0,0,1}_2)/\sqrt{2},
\end{align}
satisfy the physical-subspace, photon-number inversion, and swap symmetries.  

To show that the $\chi^{(s)}$ PCC is capable of correcting single-photon loss errors, we will test the error-correction condition for such errors under the assumption that the photon-loss probability $\gamma$ is the same for all six modes.  (This assumption entails no loss of generality as the only consequence of allowing unequal loss probabilities is the appearance of more complicated expressions in evaluating the error-correction condition.) To lowest order in $\gamma$, the Kraus operators for single-photon loss errors are~\cite{Girvin2016}:
\begin{align}\label{ErrorsLowestOrder}
&\hat{E}_0 \approx \hat{I}-\sum_{\ell =1}^6\gamma\hat{a}_{k_\ell}^\dagger\hat{a}_{k_\ell}/2,\\
&\hat{E}_\ell  \approx \sqrt{\gamma}\,\hat{a}_{k_\ell}, \, 1\le \ell \le 6,\\
&\sum_{\ell=0}^6 \hat{E}_\ell^\dagger \hat{E}_\ell = \hat{I},
\end{align}  
where $\{\hat{a}_{k_1},\hat{a}_{k_2},\ldots,\hat{a}_{k_6}\} \equiv \{\hat{a}_{s_1}, \hat{a}_{i_1}, \hat{a}_{p_1},\hat{a}_{s_2}, \hat{a}_{i_2}, \hat{a}_{p_2}\}$, and $\hat{I}$ is the identity operator.  The error-correction condition~\cite{knill1997theory} for  the single-photon loss errors associated with $\{\hat{E}_0, \hat{E}_1, \ldots, \hat{E}_6  \}$ is
\begin{align}
\bra{\tilde{a}} \hat{E}^\dagger_h \hat{E}_j\ket{\tilde{b}}=\alpha_{hj}\delta_{ab}, \,\,\,\,  \mbox{for $0\le h,j\le 6$},
\label{ECcondition}
\end{align}
where $\delta_{ab}$ is the Kronecker delta function, $\ket{\tilde{a}}, \ket{\tilde{b}}$ are arbitrary logical-basis states, and the matrix elements $\{\alpha_{hj}\}$ are independent of the $\tilde{a}$ and $\tilde{b}$ values. Equation~(\ref{ECcondition}) guarantees that no single-photon loss error distorts the code subspace, thus all of them are correctable. 

For the qubit-encoded $\chi^{(2)}$ PCC from Eqs.~(\ref{EC1qubit}) and (\ref{EC2qubit}), we find that Eq.~(\ref{ECcondition}) holds with
\begin{align}
\alpha_{00} &= \bra{\tilde{a}}(\hat{I}-\mbox{$\sum_{\ell = 1}^6$}\gamma\hat{a}_{k_\ell}^\dagger\hat{a}_{k_\ell}) \ket{\tilde{a}}= 1-3\gamma,\\
\alpha_{hh}&= \bra{\tilde{a}}\gamma \hat{a}_{k_h}^\dagger\hat{a}_{k_h} \ket{\tilde{a}}=\gamma/2, \, 1\le h \le 6,\\
\alpha_{hj}&= 0,\,\, \text{for} \, h\neq j.
\end{align}
Similarly, the qutrit-encoded $\chi^{(2)}$ PCC from Eqs.~(\ref{EC1})--(\ref{EC3}) obeys Eq.~(\ref{ECcondition}) with
\begin{align}
\alpha_{00} &= \bra{\tilde{a}}(\hat{I}-\mbox{$\sum_{\ell = 1}^6$}\gamma\hat{a}_{k_\ell}^\dagger\hat{a}_{k_\ell}) \ket{\tilde{a}}= 1-6\gamma,\\
\alpha_{hh}&= \bra{\tilde{a}}\gamma \hat{a}_{k_h}^\dagger\hat{a}_{k_h} \ket{\tilde{a}}=\gamma, \, 1\le h\le 6,\\
\alpha_{hj}&= 0,\,\, \text{for} \, h\neq j.
\end{align}
The preceding encodings both satisfy the error-correction condition because in each code the average photon number is the same for each of that code's six bosonic modes. 

The qubit and qutrit $\chi^{(2)}$ PCC's also correct single-photon gain errors. Specifically, using the photon-creation operators $\{\hat{a}_{k_h}^\dagger\}$ in lieu of the photon-loss Kraus operators $\{\hat{E}_h\}$, we satisfy the following error-correction conditions for photon-gain errors:  
\begin{equation}\nonumber
\bra{\tilde{a} }\hat{a}_{k_h} \hat{a}_{k_j}^\dagger\ket{\tilde{b} }=\delta_{hj}\delta_{ab},\, \mbox{for $a,b = 0,1$ and $1\le h,j\le 6$},
\end{equation}
for the qubit case; and
\begin{align}\nonumber
\bra{\tilde{a} }\hat{a}_{k_h} \hat{a}_{k_j}^\dagger\ket{\tilde{b} }=2\delta_{hj}\delta_{ab},\, \mbox{for $a,b = 0,1,2$ and $1\le h,j\le 6$},
\end{align}
for the qutrit case. 

At this juncture we can sketch how the photon-number parity-symmetry of each $N$-pump-photon subspace enables the $\chi^{(2)}$ PCC of dimension $N+1$ to detect single-photon (loss or gain) errors.  Let $\ket{\tilde{a}}$ be an arbitrary logical basis state of a $\chi^{(2)}$ PCC, and $\hat{n}_{k_\ell} \equiv \hat{a}_{k_\ell}^\dagger\hat{a}_{k_\ell}$ be the photon-number operator associated with the $\hat{a}_{k_\ell}$ mode.   We know that the photon-number parity vector, 
\begin{align}\nonumber
{\bf p}_{12} \equiv & \,  [\langle \hat{n}_{s_1} +\hat{n}_{i_1} \rangle, \langle\hat{n}_{s_1} +\hat{n}_{p_1}\rangle,  \langle\hat{n}_{i_1} +\hat{n}_{p_1}\rangle, \\ \label{ParityVector}
&\,\langle \hat{n}_{s_2} +\hat{n}_{i_2}\rangle, \langle \hat{n}_{s_2} +\hat{n}_{p_2}\rangle , \langle \hat{n}_{i_2} +\hat{n}_{p_2}\rangle]{\rm mod}\,2,
\end{align}
will satisfy
\begin{equation}
{\bf p}_{12} =   [2N, N, N, 2N, N, N]{\rm mod}\,2
\end{equation}
for all logical-basis states.  Assuming that loss of a single photon is the only error that has occurred, then, as shown in Appendix~\ref{ENCODE}, a nondemolition measurement of ${\bf p}_{12}$~\cite{Sun2014} yields a syndrome that identifies the mode which has lost a photon. 
If, however, the loss or gain of a single photon is the only error that has occurred, then measurement of ${\bf p}_{12}$ and the generalized photon-number parity vector, 
\begin{align}\label{q12eq}
{\bf q}_{12} \equiv [\langle\hat{n}_{s_1}+\hat{n}_{i_1}+\hat{n}_{p_1}\rangle, \langle\hat{n}_{s_2}+\hat{n}_{i_2}+\hat{n}_{p_2}\rangle]{\rm mod}\,3,
\end{align}
provides syndromes that identify which mode has suffered an error \emph{and} whether that mode lost or gained a photon.  See Appendix~\ref{ENCODE} for the details.

Thus far we have limited our attention to the qubit and qutrit $\chi^{(2)}$ PCCs.  To encode an $N$-dimensional logical qudit  in a $\chi^{(2)}$ PCC to protect against single-photon-loss  errors and single-photon-gain errors, we employ the $N$-dimensional physical-subspace symmetry operators $\{\hat{Z}^{(N)}_{s_\ell,p_\ell}, \hat{Z}^{(N)}_{i_\ell,p_\ell}: \ell = 1,2\}$ together with the photon-number inversion-symmetry operator $\otimes_{\ell=1}^2\hat{V}^{(N-1)}_\ell$ and the swap operator $\hat{X}^{(N)}_{1,2}$. For $N=2 m$ an even integer, this procedure yields the following logical-qudit basis states in terms of the three-mode physical-qudit basis states: 
\begin{widetext} 
\begin{align}
&\ket{\tilde{0}} =(\ket{m, m, m-1}_1\ket{m-1, m-1,m}_2 +\ket{m-1, m-1,m}_1\ket{m, m, m-1}_2)/\sqrt{2}, \label{PCCeven0}\\ 
&\ket{\tilde{1}}=(\ket{m, m,m-1}_1\ket{m,m,m-1}_2+\ket{m-1, m-1,m}_1\ket{m-1, m-1,m}_2)/\sqrt{2},\\
&\ket{\tilde{2}} = (\ket{m+1, m+1, m-2}_1\ket{m-2, m-2,m+1}_2+\ket{m-2,m-2,m+1}_1\ket{m+1, m+1, m-2}_2)/\sqrt{2}, \\ 
&\ket{\tilde{3}}=(\ket{m+1, m+1,m-2}_1\ket{m+1,m+1,m-2}_2+\ket{m-2, m-2,m+1}_1\ket{m-2, m-2,m+1}_2)/\sqrt{2},\\
&\vdots \nonumber \\ 
&\ket{\widetilde{N-2}} =(\ket{2m-1,2m-1,0}_1\ket{0,0,2m-1}_2+\ket{0,0,2m-1}_1\ket{2m-1,2m-1,0}_2)/\sqrt{2},\\ 
&\ket{\widetilde{N-1}}=(\ket{2m-1,2m-1,0}_1\ket{2m-1,2m-1,0}_2 +\ket{0,0,2m-1}_1\ket{0,0,2m-1}_2)/\sqrt{2}. 
\end{align}
For $N=2m+1$ an odd integer, the same approach leads to the encoding
\begin{align}
&\ket{\tilde{0}}=\ket{m,m,m}_1\ket{m,m,m}_2\\
&\ket{\tilde{1}}=(\ket{m+1, m+1, m-1}_1\ket{m+1, m+1, m-1}_2+\ket{m-1, m-1, m+1}_1\ket{m-1, m-1, m+1}_2)/\sqrt{2},\\
&\ket{\tilde{2}}=(\ket{m+1, m+1, m-1}_1\ket{m-1, m-1, m+1}_2+\ket{m-1, m-1, m+1}_1\ket{m+1, m+1, m-1}_2/\sqrt{2},\\ 
&\vdots \nonumber\\ 
&\ket{\widetilde{N-2}}==(\ket{2m, 2m, 0}_1\ket{2m, 2m, 0}_2+\ket{0, 0, 2m}_1\ket{0, 0, 2m}_2)/\sqrt{2},\\
&\ket{\widetilde{N-1}}=(\ket{2m, 2m, 0}_1\ket{0, 0, 2m}_2+\ket{0,0,2m}_1\ket{2m, 2m, 0}_2)/\sqrt{2}.\label{PCCoddN-1}
\end{align}
\end{widetext}
On average, the $\chi^{(2)}$ PCC uses $3(N-1)$ photons to encode each $N$-dimensional logical qudit into two $N$-dimensional physical qudits for protection against single-photon (loss or gain) errors.

\subsection{$\chi^{(2)}$ Embedded Error-Correcting Code}\label{chi2EECC}
The $\chi^{(2)}$ EECC encodes a single logical qudit of dimension $N$ into a single physical qudit of dimension $2N-1$ for protection against single-photon (loss or gain) errors.  Its basis states for $N$-dimensional logical qudits are the simultaneous unity-eigenvalue eigenstates of $\hat{Z}_{s,p}^{(2N-1)}$, $\hat{Z}_{i,p}^{(2N-1)}$, and $\otimes_{\ell =1}^2\hat{V}^{(2N-2)}_\ell$, i.e., they lie in $\mathcal{H}_{2N-2}$ and obey photon-number inversion symmetry.  Consider the $N =2$ case, in which a logical qubit is encoded into a physical qutrit.  Letting $|\psi\rangle = \sum_{n=0}^2 v_n|n,n,2-n\rangle$ be an arbitrary state in $\mathcal{H}_2$, we have that
$\hat{V}^{(2)}|\psi\rangle = |\psi'\rangle$, where $|\psi'\rangle = \sum_{n=0}^2 v_n'|n,n,2-n\rangle$ with 
 \begin{align}
 \left[\begin{array}{c}v'_0 \\ v'_1\\ v'_2\end{array}\right] =\begin{bmatrix}
		0 & 0&1\\
	0&1&0\\
	1&0&0 
\end{bmatrix} 
\left[\begin{array}{c}v_0 \\ v_1\\ v_2\end{array}\right].
\end{align}
Simple linear algebra then gives this relation's only unity-eigenvalue eigenstates, 
\begin{align}
&\ket{\tilde{0}}=(\ket{2,2,0}+\ket{0,0,2})/\sqrt{2},\label{encodingEECC0}\\
&\ket{\tilde{1}}=\ket{1,1,1},\label{encodingEECC1}
\end{align}
which are thus the logical-basis states for the qubit $\chi^{(2)}$ EECC. 

As is the case for the NOON code and other bosonic codes~\cite{Ludkin2014,Girvin2016}, our qubit $\chi^{(2)}$ EECC's encoding ensures that all three of its modes have the same average photon number. The principal difference from previous bosonic codes is that the physical-basis states used here span the irreducible subspace of the $\chi^{(2)}$ Hamiltonian used for universal gate constructions. This property greatly simplifies the error-correction procedure and universal transformations in the encoded basis, as will be seen in Appendix~\ref{GATE}.

Paralleling our development for the qutrit $\chi^{(2)}$ PCC, we have that, to lowest order in the photon-loss probability $\gamma$, the qubit $\chi^{(2)}$ EECC's Kraus operators for photon-loss errors are 
\begin{align}
&\hat{E}_0  \approx \hat{I} - \sum_{\ell=1}^3\gamma\hat{a}_{k_\ell}^\dagger\hat{a}_{k_\ell}/2,\\
&\hat{E}_\ell \approx \sqrt{\gamma}\,\hat{a}_{k_\ell}, 1\le \ell \le 3,\\
&\sum_{\ell=0}^3 \hat{E}_\ell^\dagger \hat{E}_\ell = \hat{I},
\end{align}  
where $\{\hat{a}_{k_1},\hat{a}_{k_2},\hat{a}_{k_3}\} \equiv \{\hat{a}_s,\hat{a}_i,\hat{a}_p\}$.  
The resulting quantum error-correction condition~\cite{knill1997theory} is therefore
\begin{align}\label{errorcorrectionconditionLoss}
\bra{\tilde{a}} \hat{E}^\dagger_h \hat{E}_j\ket{\tilde{b}}=\alpha_{hj}\delta_{ab}, \mbox{ for $0\le h,j \le 3$},
\end{align}
where $\ket{\tilde{a}}$, $\ket{\tilde{b}}$ are arbitrary logical-basis states and the $\{\alpha_{hj}\}$ are independent of the $\tilde{a}$ and $\tilde{b}$ values.  For our qubit $\chi^{(2)}$ EECC's logical-basis states, we find that Eq.~(\ref{errorcorrectionconditionLoss}) is satisfied, because direct evaluation leads to 
\begin{align}
\alpha_{00} &= \bra{\tilde{a}}(\hat{I}-\mbox{$\sum_{\ell = 1}^3$}\gamma\hat{a}_{k_\ell}^\dagger\hat{a}_{k_\ell})\ket{\tilde{a}}= 1-3\gamma,
\label{EECCcondx0}\\
\alpha_{hh}&= \bra{\tilde{a}}\gamma \hat{a}_{k_h}^\dagger\hat{a}_{k_h} \ket{\tilde{a}}=\gamma, \, 1\le h \le 3,
\label{EECCcondx1}\\
\alpha_{hj}&= 0,\,\, \text{for} \, h\neq j. \label{EECCcondx2}
\end{align}
Equations~(\ref{encodingEECC0}) and (\ref{encodingEECC1}) obey the code-space nondistortion conditions, Eqs.~(\ref{EECCcondx0})--(\ref{EECCcondx2}), because the average photon number in all three modes of each logical-basis state is identical.  Our qubit $\chi^{(2)}$ EECC also obeys the error-correction condition for single-photon gain errors, 
\begin{align}\label{errorcorrectionconditionGain}
\bra{\tilde{a}} \hat{a}_{k_h} \hat{a}_{k_j}^\dagger\ket{\tilde{b}}=2\delta_{hj}\delta_{ab},
\end{align}

Because its logical-basis states lie in $\mathcal{H}_2$, our qubit $\chi^{(2)}$ EECC's photon-number parity vector, ${\bf p}$ from Eq.~(\ref{pvectdefn}), is constant, ${\bf p} = [0,0,0]$, for all states in the code space.  Assuming that loss of a single photon is the only error that has occurred, then, as shown in Appendix~\ref{GATE}, a nondemolition measurement of ${\bf p}$~\cite{Sun2014} yields a syndrome that uniquely identifies the mode which has lost a photon.  If, however, the loss or gain of a single photon is the only error that has occurred, then measurement of ${\bf p}$ and the generalized photon-number parity,
\begin{equation}\label{genparvec}
q_{\rm EECC} \equiv [\langle \hat{n}_s + \hat{n}_i + \hat{n}_p\rangle ]{\rm mod}\,3,
\end{equation}
provides syndromes that identify which mode has suffered an error \emph{and} whether that mode lost or gained a photon.  Details of the qubit $\chi^{(2)}$ EECC's error detection and error correction appear in Appendix~\ref{GATE}. 

The $\chi^{(2)}$ EECC's logical-basis states for encoding an $N$-dimensional logical qudit into a physical qudit of $2N-1$ dimensions---found by applying the physical subspace and photon-number inversion symmetries---are easily shown to be
\begin{align}\label{chi2EECC0}
&\ket{\tilde{0}}=(\ket{2N-2,2N-2,0} + \ket{0,0,2N-2})/\sqrt{2},\\ \label{chi2EECC1}
&\ket{\tilde{1}}=(\ket{2N-3,2N-3,1} + \ket{1,1,2N-3}/\sqrt{2},\\
&\vdots \nonumber\\
&\ket{\widetilde{N-1}}=\ket{N-1,N-1,N-1}. \label{chi2EECCN}
\end{align}
Once again, each optical mode's having the same average photon number across all logical-basis states ensures that the error-correction condition for single-photon (loss or gain) errors are obeyed.  This encoding uses a total of $3(N-1)$ photons.
 
\subsection{$\chi^{(2)}$ Binomial Code}
\label{subsec:chi2bc}

Inspired by previous work~\cite{Chuang1997,Loock2016,Girvin2016}, our $\chi^{(2)}$ BC encodes a logical qubit into a $2N$-dimensional physical qudit. First, we enforce the physical-subspace symmetry characterized by $\{\hat{Z}^{(2N)}_{s ,p},\hat{Z}^{(2N)}_{i,p}\}$ to restrict the logical-basis states to the physical-qudit subspace $\mathcal{H}_{2N-1}$.  Next, to leverage binomial symmetry that will protect the code subspace from distortion by photon loss or gain errors, we introduce the symmetry described by conjugating the photon-number inversion operator $\hat{V}^{(2N-1)}$,with the pseudo-beam-splitter operator  $\hat{U}_{\rm BS}$ given by 
 \begin{align}\label{beamsplitterUnitary}
 \hat{U}_{\rm BS}\equiv \ket{\tilde{0}}\bra{+} + \ket{\tilde{1}}\bra{-} + \sum_{j=1}^{2N-2}\ket{\tilde{j}}\bra{j,j,2N-1-j},
 \end{align}
 where $\{\ket{\tilde{j}} : 0 \le j \le 2N-1\}$ is an orthonormal basis for $\mathcal{H}_{2N-1}$, 
 \begin{equation}
 \ket{\pm} \equiv (\ket{0,0,2N-1} \pm \ket{2N-1,2N-1,0})/\sqrt{2},
 \end{equation}
 and 
\begin{align}
\ket{\tilde{0}} &\equiv \sum_{j=0}^{N-1}\sqrt{\binom{2N-1}{2j}}\frac{\ket{2j,2j,2N-1-2j}}{2^{N-1}}, \label{BC0}\\
\ket{\tilde{1}} & \equiv \sum_{j=0}^{N-1}\sqrt{\binom{2N-1}{2j+1}}\frac{\ket{2j+1,2j+1,2(N-1-j)}}{2^{N-1}}
\label{BC1}
\end{align}
will soon be seen to be the $\chi^{(2)}$ BC's logical basis.  

Recall that the $\chi^{(2)}$ BC's logical-basis states are the simultaneous unity-eigenvalue eigenstates of $\hat{Z}_{s,p}^{(2N)}, \hat{Z}_{i,p}^{(2N)}$, and $\hat{\Pi}_s\hat{U}_{\rm BS}\hat{V}^{(2N-1)}\hat{U}_{\rm BS}^\dagger$.  The simultaneous unity-eigenvalue eigenstates of $\hat{Z}_{s,p}^{(2N)}$ and  $\hat{Z}_{i,p}^{(2N)}$ comprise $\mathcal{H}_{2N-1}$, so we only need concern ourselves with finding the unity-eigenvalue eigenstates of 
$\hat{\Pi}_s\hat{U}_{\rm BS} \hat{V}^{(2N-1)}\hat{U}_{\rm BS}^\dagger$. Because $2N-1$ is an odd number, $\ket{+}, \ket{-}$ are $\hat{V}^{(2N-1)}$'s only two eigenstates and their eigenvalues are 1 and $-1$, respectively.   To show that $\ket{\tilde{0}},\ket{\tilde{1}}$ from Eqs.~(\ref{BC0}) and (\ref{BC1}) are the $\chi^{(2)}$ BC's logical-basis states we first note that these states both lie in $\mathcal{H}_{2N-1}$.  Then we use our definition of $\hat{U}_{\rm BS}$ to write
\begin{eqnarray}
\lefteqn{\hat{\Pi}_s\hat{U}_{\rm BS}\hat{V}^{(2N-1)}\hat{U}_{\rm BS}^\dagger\ket{\tilde{0}} } \nonumber \\
 &=& \hat{\Pi}_s\hat{U}_{\rm BS}\hat{V}^{(2N-1)}\hat{U}_{\rm BS}^\dagger\hat{U}_{\rm BS}\ket{+} \\ 
 &=&  \hat{\Pi}_s\hat{U}_{\rm BS}\hat{V}^{(2N-1)}\ket{+} = \hat{\Pi}_s\hat{U}_{\rm BS}\ket{+} \\ 
 &=& \hat{\Pi}_s\ket{\tilde{0}} = \ket{\tilde{0}},
\end{eqnarray}
which proves that $\ket{\tilde{0}}$ is a $\chi^{(2)}$ BC logical-basis state.  A similar calculation for $\ket{\tilde{1}}$, 
\begin{eqnarray}
\lefteqn{\hat{\Pi}_s\hat{U}_{\rm BS}\hat{V}^{(2N-1)}\hat{U}_{\rm BS}^\dagger\ket{\tilde{1}} } \nonumber \\
 &=& \hat{\Pi}_s\hat{U}_{\rm BS}\hat{V}^{(2N-1)}\hat{U}_{\rm BS}^\dagger\hat{U}_{\rm BS}\ket{-} \\ 
 &=&  \hat{\Pi}_s\hat{U}_{\rm BS}\hat{V}^{(2N-1)}\ket{-} = -\hat{\Pi}_s\hat{U}_{\rm BS}\ket{-} \\ 
 &=& -\hat{\Pi}_s\ket{\tilde{1}} = \ket{\tilde{1}},
\end{eqnarray}
proves that it is the $\chi^{(2)}$ BC's other logical-basis state.

With the logical-basis states in hand, we can proceed to the error-correction conditions. The Kraus operator for there being $h$ photons lost from the signal mode, $g$ photons lost from the idler mode, and $\ell$ photons lost from the pump mode is $\hat{a}_s^h\hat{a}_i^g\hat{a}_p^\ell$  Likewise, the Kraus operator for there being $h$ photons gained by the signal mode, $g$ photons gained by the idler mode, and $\ell$ photons gained by the pump mode is $\hat{a}_s^{\dagger h}\hat{a}_i^{\dagger g}\hat{a}_p^{\dagger \ell}$.  Dephasing on mode $k = s,i,p$ caused by $\delta t$-duration dispersive propagation with dephasing rate $\dot\phi$ can be represented by the unitary operator
 \begin{align}
\hat{U}_k(\delta t)=e^{-i\delta t\,{\dot\phi}\,\hat{n}_k} 
 \approx \hat{I}_k - i\delta t\,{\dot\phi}\,\hat{n}_k  -(\delta t\,\dot\phi\,\hat{n}_k)^2/2 +\cdots,
 \end{align}
where the Taylor-series expansion shows that $m$th-order dephasing on mode $k$ has error operator $\hat{n}_k^m$, hence $\hat{n}_s^h\hat{n}_i^g\hat{n}_p^\ell$ is the error operator for $h$th-order dephasing of the signal mode, $g$th-order dephasing of the idler mode, and $\ell$th-order dephasing of the pump mode.

Suppose we are given channel-monitoring information indicating that an error of degree $m$ has occurred, i.e., $m$ photons have been lost, or $m$ photons have been gained, or $(m-1)$th-order dephasing has occurred, but there was no combination of photon-loss, photon-gain, or dephasing errors.  The relevant error-operator set is then $\xi_m=\{\hat{I} , \hat{a}_s^h\hat{a}_i^g\hat{a}_p^\ell, \hat{a}_s^{\dagger h}\hat{a}_i^{\dagger g}\hat{a}_p^{\dagger \ell },\hat{n}_s^{h}  \hat{n}_i^{g}\hat{n}_p^\ell :  h,g,\ell \ge 0, h+g+\ell = m\}$.  Using $\{\hat{E}_u^{(m)}\}$ as shorthand for the error operators in $\xi_m$, the Knill-Laflamme condition for all of these errors to be correctable is thus
\begin{equation}
\label{generalErrorCorrection}
\bra{\tilde{a}} \hat{E}_u^{(m)\dagger} \hat{E}^{(m)}_v \ket{\tilde{b}} = \alpha_{uu}\delta_{ab}\delta_{uv},
\end{equation}
for $a,b = 0,1$ and all $\hat{E}_u^{(m)},\hat{E}_v^{(m)}$ in $\xi_m$.

To verify that $\chi^{(2)}$ BC's satisfies the preceding error-correction condition. let us start with the simplest case, the $N=2$ code in which each logical qubit is encoded as
\begin{align}
&\ket{\tilde{0}}=(\ket{0,0,3}+\sqrt{3}\,\ket{2,2,1})/2,\\
&\ket{\tilde{1}}=(\ket{3,3,0}+\sqrt{3}\,\ket{1,1,2})/2.
\end{align}
It is straightforward to verify that the $N=2$ encoding satisfies the error correction condition given in Eq.~(\ref{generalErrorCorrection})  against the following error sets:
\begin{align}
\xi_0=&\{\hat{I}\}\\
\xi_1=&\{\hat{a}_s, \hat{a}_i, \hat{a}_p,\hat{a}_s^\dagger, \hat{a}_i^\dagger, \hat{a}_p^\dagger\}\\\nonumber
\xi_2=&\{\hat{a}_s\hat{a}_i, \hat{a}_s\hat{a}_p, \hat{a}_i\hat{a}_p,  \hat{a}_s^\dagger\hat{a}_i^\dagger, \hat{a}_s^\dagger\hat{a}_p^\dagger,\hat{a}_i^\dagger\hat{a}_p^\dagger,\\
& \hat{a}_s^2, \hat{a}_i^2, \hat{a}_p^2, \hat{a}_s^{\dagger 2}, \hat{a}_i^{\dagger 2}, \hat{a}_p^{\dagger 2}, \hat{n}_s, \hat{n}_i ,\hat{n}_p \}.
\end{align}  
Note that the annihilation-operator elements in $\xi_2$ correspond to discrete photon-number jumps~ \cite{liang2016cat,Girvin2016} that are not the Kraus operators commonly used for the amplitude-damping channel.  As shown in Ref.~\cite{Girvin2016}, however, correction of such discrete errors can nevertheless handle the amplitude-damping channel's lowest-order errors.

The $\chi^{(2)}$ BC with $N=3$ corrects up to three-photon-loss errors, three-photon gain errors, and second-order dephasing errors.  Its logical qubits are 
\begin{align}
&\ket{\tilde{0}}=(\ket{0,0,5} +\sqrt{10}\,\ket{2,2,3} +\sqrt{5}\,\ket{4,4,1})/4,\\
&\ket{\tilde{1}}=(\ket{5,5,0}+ \sqrt{10}\,\ket{3,3,2}+\sqrt{5}\,\ket{1,1,4})/4.
\end{align}
In addition to satisfying the error-correction conditions for $\{\xi_m: 0\le m \le 2\}$, this encoding satisfies that condition for
\begin{align}\nonumber
\xi_3=&\{  \hat{a}_s\hat{a}_i  \hat{a}_p,  \hat{a}_s^\dagger\hat{a}^\dagger_i  \hat{a}^\dagger_p, \hat{a}_s^2\hat{a}_i,  \hat{a}_s^2\hat{a}_p, \hat{a}_s\hat{a}_i^2, \hat{a}_s\hat{a}_p^2,\hat{a}_i^2\hat{a}_p, \hat{a}_i\hat{a}_p^2, \\\nonumber
&  \hat{a}_s^{\dagger 2}\hat{a}_i^\dagger,  \hat{a}_s^{\dagger 2}\hat{a}_p^\dagger, \hat{a}_s^\dagger\hat{a}_i^{\dagger 2}, \hat{a}_s^\dagger\hat{a}_p^{\dagger 2},\hat{a}_i^{\dagger2}\hat{a}^\dagger_p,\hat{a}_i^\dagger\hat{a}_p^{\dagger 2},\hat{a}_s^3, \hat{a}_i^3, \hat{a}_p^3, \\
& \hat{a}_s^{\dagger 3}, \hat{a}_i^{\dagger 3}, \hat{a}_p^{\dagger 3}, \hat{n}_s\hat{n}_i, \hat{n}_s\hat{n}_p, \hat{n}_i\hat{n}_p, \hat{n}_s^2, \hat{n}_i^2, \hat{n}_p^2\}.
\end{align}  
In general, the $\chi^{(2)}$ BC on $\mathcal{H}_{2N-1}$ protects a logical qubit against errors in $\{\xi_m : 1\le m \le N\}$, by means of the encoding from Eqs.~(\ref{BC0}) and (\ref{BC1}).

Let us now verify that the $\chi^{(2)}$ BC's encoding, Eqs.~(\ref{BC0}) and (\ref{BC1}), satisfies the error-correction conditions for $\{\xi_m : 0\le m \le N\}$, starting with the orthogonality condition. Orthogonality here means that any correctable error applied to the logical-basis states $\ket{\tilde{0}}$ and $\ket{\tilde{1}}$ will result in orthogonal states.  For photon-loss or photon-gain errors, orthogonality is satisfied by the $\chi^{(2)}$ BC because the photon-number parities of the signal, idler, and pump modes in its $\ket{\tilde{0}}$ state are opposite those of its $\ket{\tilde{1}}$ state.  Consequently, photon-loss or photon-gain errors of order $m\leq N$ do not disturb orthogonality, because any such error's modal-parity flips are the same for $\ket{\tilde{0}}$ and  $\ket{\tilde{1}}$.  

The $\chi^{(2)}$ BC's encoding also leads to orthogonality between the error syndromes---obtained from photon-number parity measurements---for photon-loss and photon-gain errors. For such an error to transform one physical-basis state to another requires pump-mode photon losses (gains) to be matched by identical gains (losses) in the signal and idler modes.  As a result,  photon-loss or photon-gain errors of order $m$ acting on a logical-basis state lead to orthogonal error syndromes.  Note, however, that the error syndromes for different dephasing errors are not orthogonal.  But, because all dephasing errors can be corrected by projecting the state onto the code subspace~\cite{Girvin2016}, orthogonality is not required for dephasing errors to be correctable.

Now let us turn to verifying the nondistortion condition in Eq.~(\ref{generalErrorCorrection}) for the $m$th-order ($m \le N$) photon-loss error, represented by the error operator $\hat{E}_u \equiv \hat{a}_s^h\hat{a}_i^g\hat{a}_p^{m-g-h}$. Without loss of generality we will assume $g\geq h$, owing to the symmetry between signal and idler modes. Using this error operator, together with Eqs.~(\ref{BC0}) and (\ref{BC1}), in Eq.~(\ref{generalErrorCorrection}) we  get:
\begin{widetext}
\begin{align}\nonumber
\bra{\tilde{0}}\hat{E}_u^\dagger\hat{E}_u\ket{\tilde{0}}&= \bra{\tilde{0}} \prod_{\ell_1=0}^{h-1}(\hat{n}_s-\ell_1)\prod_{\ell_2=0}^{g-1}(\hat{n}_i-\ell_2)\prod_{\ell_3=0}^{m-h-g-1}(\hat{n}_p-\ell_3)\ket{\tilde{0}},\\\nonumber
&=\frac{1}{2^{2N-2}}\sum_{j=\lceil \frac{g}{2}\rceil}^{\lfloor N-\frac{m-g-h+1}{2}\rfloor} \binom{2N-1}{2j}\prod_{\ell_1=0}^{h-1}(2j-\ell_1)\prod_{\ell_2=0}^{g-1}(2j-\ell_2)\prod_{\ell_3=0}^{m-g-h-1}(2N-2j-1-\ell_3),\\\label{ZERO}
&=\frac{1}{2^{2N-2}}\sum_{j=\lceil \frac{g}{2}\rceil}^{\lfloor N-\frac{m-g-h+1}{2}\rfloor}\frac{(2N-1)!(2j)!}{(2j-h)!(2j-g)!(2N-2j-1-m+h+g)!},
\end{align}
\begin{align}\nonumber
\bra{\tilde{1}}  \hat{E}_u^\dagger \hat{E}_u\ket{\tilde{1}}&=\bra{\tilde{1}}\prod_{\ell_1=0}^{h-1}(\hat{n}_s-\ell_1)\prod_{\ell_2=0}^{g-1}(\hat{n}_i-\ell_2)\prod_{\ell_3=0}^{m-h-g-1}(\hat{n}_p-\ell_3)\ket{\tilde{1}},\\\nonumber
&=\frac{1}{2^{2N-2}}\sum_{j^\prime=\lceil \frac{m-h-g}{2}\rceil}^{\lfloor N-\frac{g+1}{2}\rfloor} \binom{2N-1}{2j} \prod_{\ell_1=0}^{h-1}(2N-2j^\prime-1-\ell_1)\prod_{\ell_2=0}^{g-1}(2N-2j^\prime-1 -\ell_2)\prod_{\ell_3=0}^{m-g-h-1}(2j^\prime-\ell_3),\\ \label{One}
&=  \frac{1}{2^{2N-2}}\sum_{j^\prime=\lceil \frac{m-h-g}{2}\rceil}^{\lfloor N-\frac{g+1}{2}\rfloor}\frac{(2N-1)!(2N-2j^\prime-1)!}{(2N-2j^\prime-1-h)!(2N-2j^\prime-1-g)!(2j^\prime-m+h+g)!}.
\end{align}
\end{widetext}
Under the change of variable $j=(N-1)/2-j^\prime$, it is straightforward to see that the right-hand sides of Eqs.~(\ref{ZERO}) and (\ref{One}) are equal. Because this result applies for all $m\le N$, and because we have already shown orthogonality, we have that the encoding in Eqs.~(\ref{BC0}) and (\ref{BC1}) satisfies the error-correction condition for $m\le N$ photon-loss errors.
    
Given channel-monitoring information indicating that an $m$th-order photon-loss error has occurred, we can identify the exact type of that error by measuring the photon-number parity vector~\cite{Mirrahimi2014} 
\begin{equation}
{\bf p}_{\rm BC} \equiv [\langle \hat{n}_s-\hat{n}_i\rangle, \langle \hat{n}_s+\hat{n}_p\rangle, \langle \hat{n}_i+\hat{n}_p\rangle]{\rm mod }(2N-1).
\end{equation}
If, however, that monitoring does not distinguish between photon-loss and photon-gain errors, then we also need to measure the generalized photon-number parity,
\begin{equation}
q_{\rm BC} \equiv [\langle \hat{n}_s + \hat{n}_i + \hat{n}_p\rangle]{\rm mod}\,(6N-3),
\end{equation}
to know whether the $m$th-order error that occurred was a loss error or a gain error.  Assuming it was an $m$th-order ($m\le N$) photon-loss error,  the number of configurations for distributing the loss of $m$ photons across the signal, idler, and pump modes is
\begin{align}
\sum_{h=0}^m\sum_{g=0}^{m-h}1 =\frac{(m+2)(m+1)}{2}\leq \frac{(N+2)(N+1)}{2}.
\end{align}
For $N\ge 2$, this number of configurations is less than the $(2N-1)^3$ possible parity vectors ${\bf p}_{\rm BC}$. Thus, for $m\le N$ our parity-measurement scheme uniquely identifies the photon-loss error that has occurred from the error set $\xi_m$.  

Now consider the $m$th-order ($m\le N$) photon-gain error, represented by the error operator $\hat{E}_u \equiv \hat{a}_s^{\dagger h}\hat{a}_i^{\dagger g}\hat{a}_p^{\dagger m-g-h}$.  Again, without loss of generality, we will presume  $g \geq h$, because of the symmetry between signal and idler modes. The nondistortion condition for the $m$th-order ($m\le N$) photon-gain error is guaranteed by the equality between the following two terms:
\begin{widetext}
\begin{align}\nonumber
\bra{\tilde{0}}  \hat{E}_u^\dagger \hat{E}_u\ket{\tilde{0}}&= \bra{\tilde{0}} \prod_{\ell_1=1}^{h}(\hat{n}_s+\ell_1)\prod_{\ell_2=1}^{g}(\hat{n}_i+\ell_2)\prod_{\ell_3=1}^{k-h-g}(\hat{n}_p+\ell_3)\ket{\tilde{0}}\\\nonumber
&=\frac{1}{2^{2N-2}}\sum_{j=0}^{N-1} \binom{2N-1}{2j}\prod_{\ell_1=1}^{h}(2j+\ell_1)\prod_{\ell_2=1}^{g}(2j+\ell_2)\prod_{\ell_3=1}^{k-g-h}(2N-2j-1+\ell_3),\\\label{ZEROgain}
&=\frac{1}{2^{2N-2}}\sum_{j=0}^{N-1} \binom{2N-1}{2j}\frac{(2j+g)!(2j+h)!(2N-2j-1+k-g-h)!}{(2j)!(2j)!(2N-2j-1)!}
\end{align}
\begin{align}
\nonumber
\bra{\tilde{1}}  \hat{E}_u^\dagger \hat{E}_u\ket{\tilde{1}}&=\bra{\tilde{1}}\prod_{\ell_1=1}^{h}(\hat{n}_s+\ell_1)\prod_{\ell_2=1}^{g}(\hat{n}_i + \ell_2)\prod_{\ell_3=1}^{k-h-g}(\hat{n}_p+\ell_3)\ket{\tilde{1}}\\\nonumber
&=\frac{1}{2^{2N-2}}\sum_{j^\prime= 0}^{N-1 }\binom{2N-1}{2j^\prime} \prod_{\ell_1=1}^{h}(2N-2j^\prime-1+\ell_1)\prod_{\ell_2=1}^{g}(2N-2j^\prime-1+\ell_2)\prod_{\ell_3=1}^{k-g-h}( 2j^\prime+\ell_3),\\ \nonumber
&=  \frac{1}{2^{2N-2}}\sum_{j^\prime= 0}^{N-1 }\binom{2N-1}{2j^\prime }\frac{(2N-2j^\prime-1+g)!(2N-2j^\prime-1+h)!(2j^\prime + k-g-h)!}{(2N-2j^\prime-1)!(2N-2j^\prime-1)!(2j^\prime)!}\\\label{Onegain}
&=  \frac{1}{2^{2N-2}}\sum_{j^\prime= 0}^{N -1}\binom{2N-1}{2N-2j-1}\frac{(2j+g)!(2j+h)!(2N-2j-1+k-g-h)!}{(2j)!(2j)!(2N-2j-1)!},
\end{align}
\end{widetext}
where the last line employed the change of variable $j=(N-1)/2-j^\prime$. Because the binomial coefficients in  Eqs.~(\ref{ZEROgain}) and (\ref{Onegain}) coincide, we see that $\bra{\tilde{0}}  \hat{E}_u^\dagger \hat{E}_u\ket{\tilde{0}} = \bra{\tilde{1}}  \hat{E}_u^\dagger \hat{E}_u\ket{\tilde{1}}$.  Combined with the previously shown orthogonality condition, we have that encoding in Eqs.~(\ref{BC0}) and (\ref{BC1}) satisfies the error-correction condition $m\le N$-photon-gain errors.   Moreover, as alluded to earlier, given channel-monitoring information indicating that an $m$th-order photon-gain error has occurred, the exact type of that error is revealed by measuring the photon-number parity vector ${\bf p}_{\rm BC}$.  If, however, that monitoring does not distinguish between photon-gain and photon-loss errors, then, as was the case earlier, we also need to measure the generalized photon-number parity, $q_{\rm BC}$, to know whether the $m$th-order error that occurred was a gain error or a loss error.   
 
Finally, we demonstrate the error-correction condition for any $m$th-order ($m+1\le N$) dephasing error $\hat{n}_s^{h}\hat{n}_i^{g}\hat{n}_p^{m-g-h-1}$.  We have that
\begin{widetext}
\begin{align}\label{phaseerror1}
\bra{\tilde{0}}\hat{n}_s^{2h}  \hat{n}_i^{2g}\hat{n}_p^{2(m-g-h)}\ket{\tilde{0}}=\frac{1}{2^{2N-2}}\sum_{j=0}^{N-1}\binom{2N-1}{2j}(2j)^{2(g+h)}(2N-2j-1)^{2(m-g-h)},\\\label{phaseerror2}
\bra{\tilde{1}}\hat{n}_s^{2h}  \hat{n}_i^{2g}\hat{n}_p^{2(m-g-h)}\ket{\tilde{1}}=\frac{1}{2^{2N-2}}\sum_{j^\prime=0}^{N-1} \binom{2N-1}{2N-2j^\prime-1}(2j^\prime-1)^{2(g+h)}(2j^\prime)^{2(m-g-h)}.
\end{align}
\end{widetext}
Making the change of variable $j=(N-1)/2-j^\prime$ shows that Eqs.~(\ref{phaseerror1}) and~(\ref{phaseerror2}) agree, thus the error-correction condition is satisfied.  The encoding, decoding, error-correction and universal logical-basis gate sets for the $\chi^{(2)}$ BC are all realizable with linear optics and $\chi^{(2)}$ Hamiltonian evolutions, but their detailed construction is beyond the scope of the current work.  

Now let us return to the issue---raised briefly earlier---of the $\chi^{(2)}$ BC's behavior with respect to the exact Kraus operators for the amplitude-damping channel's $m$-photon loss error on the $\ell$th bosonic mode.  These Kraus operators are~\cite{Chuang1995}:
\begin{eqnarray}
\lefteqn{ \hat{A}_\ell(m) 
  =\sqrt{\frac{\gamma^m}{m!}}\,(1-\gamma)^{\hat{a}_\ell^\dagger\hat{a}_\ell/2}\hat{a}_\ell^m,}\\
  &=&\sum_{n=m}^\infty \sqrt{\binom{n}{m}}\sqrt{\gamma^m}\sqrt{(1-\gamma)^{n-m}}\,\ket{n-m}\bra{n},
\end{eqnarray}
and they satisfy $\sum_{m=0}^\infty \hat{A}^\dagger_\ell(m)\hat{A}_\ell(m)=\hat{I}_\ell $. The factor of $(1-\gamma)^{\hat{a}_\ell^\dagger\hat{a}_\ell/2}$ in $\hat{A}_\ell(m)$ implies that all bosonic modes must share a common photon-number sum if $m$-photon loss errors are to be correctable, and our $\chi^{(2)}$ BC's encoding, Eqs.~(\ref{BC0}) and (\ref{BC1}), fails to obey this condition

To circumvent the preceding difficulty with the amplitude-damping channel,  we generalize our three-mode encoding to  the following  two-mode encoding:
\begin{align} \label{newEncoding}
\ket{\tilde{0^\prime}}&= \frac{1}{2^{N-1}}\sum_{j=1}^N\sqrt{\binom{2N-1}{2j-1}}\ket{ 2j, 2N-2j-1}, \\\label{NewEncoding2}
\ket{\tilde{1^\prime}}
&=\frac{1}{2^{N-1}}\sum_{j=0}^{N-1}\sqrt{\binom{2N-1}{2j}}\ket{2N-2j-1,2j},
\end{align} 
whose physical-basis states are $\{\ket{n_s,n_p} : 0\le n_s,n_p; n_s+n_p = 2N-1\}$ with $\ket{n_s,n_p}$ denoting a Fock state containing $n_s$ signal photons and $n_p$ pump photons.  These physical-basis states no longer lie in an irreducible subspace of the $\chi^{(2)}$ Hamiltonian, hence they cannot be prepared with just linear optics and $\chi^{(2)}$ Hamiltonian evolutions as is the case for our other $\chi^{(2)}$ QEC codes.  They are stabilized instead by the symmetry operators  $\hat{Z}_{1,2}^{(2N+1)}$ and $\hat{\Pi}_s\hat{\tilde{U}}_{\rm BS}\hat{\tilde{V}}^{(2N)}\hat{\tilde{U}}_{BS}^{\prime\dagger}$, where 
$\hat{\tilde{V}}^{(2N)}$ and $\hat{\tilde{U}}_{\rm BS}$ and  are two-mode generalizations of Eqs.~(\ref{Voperator}) and (\ref{beamsplitterUnitary}) obtained by treating signal and idler as a single mode.

This two-mode encoding might be preparable in a hybrid system that combines $\chi^{(2)}$ interactions with Jaynes-Cumming interactions or four-wave mixing, such as can be realized in superconducting resonators~\cite{Jiang2015A,Jiang2015B}.  More importantly, for our present purpose, the two-mode encoding obeys the error-correction condition 
\begin{equation}
\bra{\tilde{a}}\hat{E}^\dagger_{h(m-h)}(m)\hat{E}^\dagger_{g(g-h)}(m)\ket{\tilde{b}} = \alpha_{hh}\delta_{ab}\delta_{hg}
\end{equation}
for $a,b = 0,1$, $0\le h,g \le m$, and $m \le N$, where $\hat{E}_{hk}(m) \equiv \hat{A}_s(h)\hat{A}_p(m-h)$.  
So, given channel monitoring that identifies the occurrence of an $m$th-order ($m \le N$) photon loss error produced by the amplitude-damping channel, the two-mode encoding in Eqs.~(\ref{newEncoding}) and (\ref{NewEncoding2}) can correct that error.  This capability derives from the photon-number sum of the signal and idler modes being the same for the two logical-basis states. 

Compared to the binomial code proposed in Ref.~\cite{Girvin2016}, our encodings in Eqs.~(\ref{BC0}), (\ref{BC1}) and  Eqs.~(\ref{newEncoding}), (\ref{NewEncoding2}) have a constant photon-number spacing in their physical-basis states, instead of the linearly growing photon-number spacing  in  Ref.~\cite{Girvin2016}. 
Also, we require only a constant number of bosonic modes to correct  $N$th-order photon-loss errors, instead of the $O(N)$ bosonic modes used by the NOON code for this purpose \cite{Loock2016}.  As a result, our binomial codes need on average $3(N-1/2)$ photons to encode a logical qubit in a manner capable of correcting $N$th-order photon-loss errors, whereas  $O(N^2)$ photons are required for other  QEC codes that have this error-correction power~\cite{Chuang1995,Gilchrist2005,Ludkin2014,Loock2016,Girvin2016,liang2016cat}. 
This advantage arises because our encodings are designed to work with channel monitoring that identifies the error order, while codes that use many more photons handle all $m\le N$ orders without any such monitoring. 

The physical fault-tolerance of our $\chi^{(2)}$  QEC codes is based on the low likelihood of the environment inducing a $\chi^{(2)}$-Hamiltonian evolution, something that is necessary to create a logical error.  That said, we have yet to consider over/under-rotation errors in the gate implementation itself. Thus a full treatment of our $\chi^{(2)}$  QEC codes' fault tolerance remains to be supplied.


\section{Universal Gate Sets in the Encoded Basis}
\label{sec:univencod}

Our bosonic codes attain their hardware efficiency by virtue of acting within a carefully crafted subspace, with well-defined symmetry properties.  A notable cost of employing such protective symmetries is a substantial rise in the number of primitive operations needed to realize logical gates on the encoded states for quantum computation.  Moreover, the coherent $\chi^{(2)}$ interaction Hamiltonian by itself is clearly not universal on all bosonic quantum states, at the least because of its symmetries.  Thus, it is important to consider: how universal quantum computation can be achieved, in principle, on encoded states of our $\chi^{(2)}$ QEC codes; what the implementation cost is to realize basic logical-gate primitives such as the controlled-NOT operation; and what interaction and control Hamiltonians are needed to attain universality.

Below, in this section, we construct and elaborate on universal gate sets for the qutrit $\chi^{(2)}$ PCC (in Sec.~\ref{subsec:univpcc}) and the qubit $\chi^{(2)}$ EECC (in Sec.~\ref{subsec:univeecc}).  We find that it is sufficient to solely employ coherent $\chi^{(2)}$ interactions and linear optics (e.g., phase shifters and beam splitters).  And although these constructions are not fault-tolerant, our explicit circuits give practical lower bounds on the complexity required for logical operations on the encoded states.  

\subsection{Qutrit $\chi^{(2)}$ Parity-Check Code}
\label{subsec:univpcc}

The qutrit $\chi^{(2)}$ PCC, defined in Eqs.~(\ref{EC1})--(\ref{EC3}), encodes each logical qutrit into two physical qutrits. Since  $\chi^{(2)}$ interactions and linear optics are sufficient for universal computation in the physical-qutrit basis~\cite{Niu2017},  they are also universal in the logical qutrit basis supported in the two-qutrit subspace
\begin{align}
\mathcal{H}_2^{\otimes 2}&=\text{Span}\{\ket{0,0,2}_1\ket{0,0,2}_2,\ket{0,0,2}_1\ket{1,1,1}_2, \nonumber \\
& \ket{0,0,2}_1\ket{2,2,0}_2,\ldots,\ket{2,2,0}_1\ket{0,0,2}_2, \nonumber \\ 
&  \ket{2,2,0}_1\ket{1,1,1}_2, \ket{2,2,0}_2 \ket{2,2,0}_2\}.
\end{align} 
We specify the detailed construction of  CZ gate below using a quantum Fredkin  gate defined as:  
\begin{widetext}
\begin{align}
\hat{F}= (\ket{0,0,2}_{1\,1}\bra{0,0,2} + \ket{1,1,1}_{1\,1}\bra{1,1,1})\otimes \hat{I}_2 + \ket{2,2,0}_{1\,1}\bra{2,2,0}\otimes (\ket{2,2,0}_{2\,2}\bra{0,0,2} + \ket{0,0,2}_{2\,2}\bra{2,2,0}),
\label{CNOTfirst}
\end{align}
\end{widetext}
which realizes a controlled swap between the $\ket{0,0,2}_2$ and $\ket{2,2,0}_2$ states conditioned on the first qutrit being in the state $ \ket{2,2,0}_1 $. In the encoded basis,  Eqs.~(\ref{EC1})--(\ref{EC3}), the CZ gate can now be realized as follows.   First, apply the quantum Fredkin gate from Eq.~(\ref{CNOTfirst}) separately to the encoded control and target states to transform each of their logical-qutrit basis states into:
\begin{align}
\label{EC1prime}
&\ket{\tilde{2}^\prime }=(\ket{2,2,0}_1  +\ket{0,0,2}_1)\ket{2,2,0}_2/\sqrt{2},\\
&\ket{\tilde{1}^\prime }=(\ket{2,2,0}_1  +\ket{0,0,2}_1)\ket{0,0,2}_2/\sqrt{2},\\\label{EC3prime}
&\ket{\tilde{0}^\prime }=\ket{1,1,1}_1\ket{1,1,1}_2.
\end{align}
Next, because the logical-basis states' second qutrits are in the computational basis, we apply a physical-basis CZ gate to the second qutrit of the control and target's encoded states.  Finally we apply the adjoint of the quantum Fredkin gate from Eq.~(\ref{CNOTfirst}) separately to the control and target states to return their logical-qutrit basis states to those from Eqs.~(\ref{EC1})--(\ref{EC3}).  The overall operation then realizes ${\rm CZ}_{c,t}$, the CZ gate between the control and target qutrit's logical-basis states, as follows
\begin{equation}
{\rm CZ}_{c,t} = (\hat{F}_c\otimes \hat{F}_t)^\dagger {\rm CZ}_{2,2}(\hat{F}_c\otimes \hat{F}_t),
\end{equation}
where $\hat{F}_k$ denotes the Eq.~(\ref{CNOTfirst}) gate applied to the control ($k=c$) or target ($k=t$) states, and ${\rm CZ}_{2,2}$ denotes the CZ gate in the qutrit basis between  the second physical-qutrit of the control and target's encoded states.

\subsection{Qubit $\chi^{(2)}$ Embedded Error-Correcting Code}
\label{subsec:univeecc}

Using only $\chi^{(2)}$ interactions and linear optics, we now show how to construct the strictly universal gate set for the qubit $\chi^{(2)}$ EECC's logical-basis states that consists of the controlled phase gate $\Lambda(S)$ and the Hadamard gate $\hat{H}$.  First we introduce the $\hat{X}_P$ gate, which rotates the logical-basis state in Eq.~(\ref{encodingEECC0}) back to a three-mode Fock state while leaving the three-mode Fock state in Eq.~(\ref{encodingEECC1}) unchanged:
\begin{align}
\hat{X}_P &= \ket{2,2,0}\bra{\tilde{0}} + \ket{1,1,1}\bra{\tilde{1}} \nonumber \\
&\quad + \ket{0,0,2}(\bra{0,0,2} - \bra{2,2,0})/\sqrt{2}.
\label{XP}
\end{align}
The $\hat{X}_P$ gate is realizable with $\chi^{(2)}$ interactions~(see Appendix~\ref{GATE}), and it serves as a computational primitive for implementing the qubit $\chi^{(2)}$ EECC's $\Lambda(S)$ gate and its Hadamard gate, as well as its encoding and error-correction operations, as explained in Appendix~\ref{GATE}. 

The Hadamard gate in the encoded qubit basis corresponds to the transformation:
\begin{align}
\hat{H}  &=[(\ket{2,2,0}+\ket{0,0,2})/2 
+\ket{1,1,1}/\sqrt{2}]\bra{\tilde{0}} \nonumber \\
&\quad+ [(\ket{2,2,0}+\ket{0,0,2})/2 - \ket{1,1,1}/\sqrt{2}]\bra{\tilde{1}} \nonumber \\
&\quad + (\ket{0,0,2}-\ket{2,2,0})(\bra{0,0,2}-\bra{2,2,0})/2.
\label{Had}
\end{align}
Using $\hat{X}_P$ gate and its inverse, we can rewrite the Hadamard gate as
$\hat{H}= \hat{X}_P^{-1}\hat{H}^\prime\hat{X}_P$,
where 
\begin{align}
&\hat{H}^\prime\ket{1,1,1}=(\ket{2,2,0}-\ket{1,1,1})\bra{1,1,1}/\sqrt{2}, \nonumber \\
&\quad +(\ket{2,2,0}+\ket{1,1,1})\bra{2,2,0}/\sqrt{2} + \ket{0,0,2}\bra{0,0,2},
\label{Hadprime}
\end{align}
is the Hadamard gate in the $\{\ket{2,2,0}, \ket{1,1,1}\}$ qubit  basis. We show in Appendix~\ref{GATE} that $\hat{H}^\prime $ can also be implemented with just $\chi^{(2)}$ interactions. 

\begin{figure}[h]
\begin{center}
\includegraphics[width=1\linewidth]{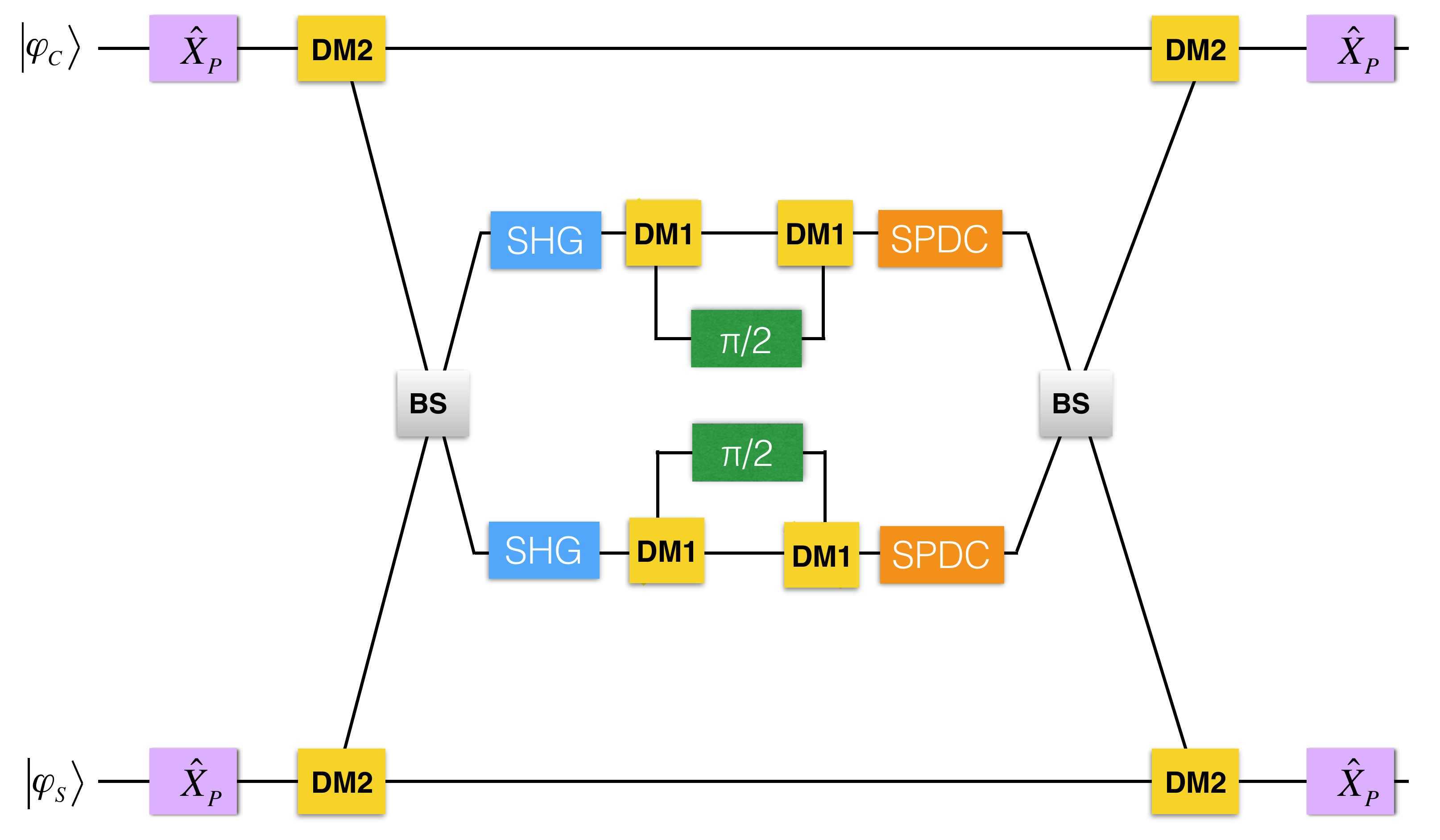}
\caption{$\Lambda(S)$ gate implementation for the qubit $\chi^{(2)}$ EECC in the logical basis.  $\ket{\phi_C}$ and $\ket{\phi_S}$:  control and target qubits.  DM2, DM1: dichroic mirrors.  BS: beam splitters.  SHG:  second-harmonic generation.  $\pi/2$:  quarter-wave phase shifter.  SPDC: type-I phase-matched spontaneous parametric downconversion. 
\label{CSGATeencoded}}
\end{center}
\end{figure}
To complete the universal gate set in the logical basis, we can implement the controlled phase gate $\Lambda(S)$ with the optical circuit shown in Fig.~\ref{CSGATeencoded}, in which $\ket{\phi}_c$ and $\ket{\phi}_t$ represent the control and target qubit's logical-basis states~\cite{footnote}.  In this circuit, the initial $\hat{X}_P$ gates rotate those logical qubits' bases back to Fock states.  Then the first set of DM2 dichroic mirrors direct the pump-mode photons into the first beam-splitter~(BS), while leaving the signal and idler photons propagating on their original rails toward the second set of DM2 dichroic mirrors.  If and only if the control and target qubits are in their $\ket{\tilde{1}}$ states does Hong-Ou-Mandel interference occur at the first BS block.  When that interference occurs, illumination of a subsequent SHG block by a frequency-$\omega_p$ two-photon Fock state---if present---results in its conversion to a frequency-$2\omega_p$ single-photon Fock state that is directed (by a DM1 dichroic mirror) to a wave plate that imparts a $\pi/2$ phase shift.  The remaining BS, DM, and (type-I phase-matched) SPDC stages complete the  $\Lambda(S)$ gate by restoring the pump-photon frequencies on the target and control rail's $\ket{\tilde{1}}$ states to $\omega_p$.  The $\Lambda(S)$ gate is completed by the final $\hat{X}_P$ gates that rotate the logical qubits' bases back to the $\chi^{(2)}$ EECC's $\{\ket{\tilde{0}},\ket{\tilde{1}}\}$.  


\section{Generalized Quantum Hamming Bounds}
\label{hamming}

New quantum Hamming bounds are essential for establishing the code-rate optimalities of our $\chi^{(2)}$ PCC and $\chi^{(2)}$ EECC, because the conventional quantum Hamming bound presumes that the physical and logical bases have the same dimensionality, whereas such is not the case for our $\chi^{(2)}$ PCC and $\chi^{(2)}$ EECC.  Furthermore, the dominant errors for our codes are photon-loss errors, not physical-qudit rotation errors.  Thus in this section we develop generalized quantum Hamming bounds to account for both of these discrepancies.  

First, in Sec.~\ref{quditrotationerrors}, we establish the generalized quantum Hamming bound for physical-qudit rotation errors.  Then, in Sec.~\ref{subsec:hammloss}, we derive the generalized quantum Hamming bound for an $[[n\log_2(q),k\log_2(b),3]]$ code that corrects all single-photon loss errors, and show that our the $\chi^{(2)}$ PCC and the $\chi^{(2)}$ EECC saturate this bound with $N=2$.
 
\subsection{Generalized Quantum Hamming Bound for Qudit-Rotation Errors\label{quditrotationerrors}} 

The generalized quantum Hamming bound for qudit-rotation errors is established in the following theorem.\\  

\noindent \textbf{Theorem 1.} The $[[n\log_2(q),k\log_2(b),  2t +1]]$ code has quantum Hamming bound given by:
\begin{align}\label{QHB}
\sum_{j=0}^t \left(\begin{array}{c}
n\\
j
\end{array}\right) (q^2-1)^jb^k \leq q^n.
\end{align}

\noindent \textit{Proof:} Suppose that there are $j\le t$ physical qudits with errors.  Their  
$\binom{n}{j}$ possible locations within the length-$n$ codeword can be determined completely by that weight-$j$ error's decomposition into $(q^2-1)^j$  independent error operators.  The dimensionality of all weight-$j$ errors for each logical codeword is therefore $\binom{n}{j}(q^2-1)^j$. So, because there are $b^k$ codewords, the total number of possible errors is 
\begin{align}
\sum_{j=0}^t\binom{n}{j}(q^2-1)^jb^k, \nonumber
\end{align}
and for them to be correctable that total should no larger than the subspace dimensionality for $n$ physical qudits of dimension $q$, i.e., we require
\begin{align}\label{GQHB}
\sum_{j=0}^t\binom{n}{j}(q^2-1)^jb^k \leq q^n,
\end{align}
which completes the proof.

Our generalized quantum Hamming bound reduces to the conventional Hamming bound for qubit encoding of qubits codes by choosing  $b=q=2 $. The advantage of adopting different dimensions for the logical and physical bases can now be illustrated.  If we use $n$ physical \textit{qutrits} to protect one logical \textit{qubit} against physical qutrit-rotation errors Theorem~1 implies that
\begin{align}
2(1+8n)\leq 3^n,
\label{GQHB2}
\end{align}
which is satisfied by $n\ge 4$, which is less than the $n=5$ required for encoding a logical qubit in the qubit basis. This example makes it natural to ask what is the maximum $k/n$ for either the same-basis or different-bases encoding.  Theorems 2 and 3 answer this question for $k=t=1.$\\

\noindent\textbf{Theorem 2.} For nondegenerate $[[n\log_2(q),\log_2(q), 3]]$ QEC codes, $\max_{q} (1/n)=1/4$ is achieved for $q \ge 4$.\\

\noindent\textit{Proof:} When $b=q$ and $k=t=1$ our generalized quantum Hamming bound from Eq.~(\ref{QHB}) simplifies to
\begin{equation}
1+n(q^2-1)\leq q^{n-1}.
\end{equation}
For $n=2$ and $n=3$ and all $q \ge 2$, this condition is never satisfied, but for $n=4$ it is satisfied for all $q\ge 4$, and the theorem is proved.  \\ 

\noindent\textbf{Theorem 3.} For nondegenerate $[[n\log_2(q), \log_2(b), 3]]$ QEC codes, $\max_b(1/n) = 1/3$ is achieved with $b=2$ for all $q\ge 6$. \\

\noindent\textit{Proof:} When $k=t=1$ our generalized quantum Hamming bound becomes
\begin{equation}
[1+n(q^2-1)]b\leq q^n.
\label{theorem3QHB}
\end{equation}
For $n=2$ this condition is equivalent to $(2b-1)q^2 -1 \le 0$, which cannot be satisfied for $b\ge 2$ and $q\ge 2$.  Direct evaluation of (\ref{theorem3QHB}) for $b=2$, however, verifies that it is violated for $2\le q \le 5$, but satisfied for $q\ge 6$, hence completing the proof. \\ 

The preceding theorems address code efficiency for situations in which the physical qudits have unlimited dimensionality but their number is fixed.  In this case, using  higher-dimensional physical qudits  to encode lower-dimensional logical qudit is advantageous. On the other hand, if codeword dimensionality, $q^n$, is fixed, a more appropriate code-efficiency metric for a QEC code is its volume ratio, $r\equiv b^k /q^n$. Theorem~4 shows that $b=q=2$ is optimum for maximizing $r$. \\  

\noindent\textbf{Theorem 4.} For nondegenerate $ [[n\log_2(q), k\log_2(b),3]] $ QEC codes, $\max_{b,q}(r)$ is attained at $b=q=2$ for all values of $k$.\\ 

\noindent\textit{Proof:} When $t=1$ our generalized quantum Hamming bound is
\begin{equation}\label{theorem4}
[1 + n(q^2-1)]b^k \le q^n,
\end{equation}
which immediately gives us
\begin{equation}
r \le 1/[1+n(q^2-1)].
\label{rOpt}
\end{equation}
The right-hand side of this inequality is maximized by $q=2$.  For $q=2$ and any $k$, choosing $b=2$ then minimizes the $n$ value needed to satisfy (\ref{theorem4}), hence maximizing $r$.  Indeed, ignoring the integer constraint on $k$, we have that $k=n-\ln(1+3n)$ achieves $r=1/(1+3n)$ when $b=q=2$.\\

Theorem~4 shows the inherent volume-efficiency advantage of choosing the physical and logical bases to be qubit bases when physical-qudit rotation errors are the errors of interest.   The situation is different, however, when qudit-rotation errors are much less likely to occur than photon-loss errors, as we will now show.


\subsection{Generalized Quantum Hamming Bounds for Photon-Loss Errors}
\label{subsec:hammloss}

The generalized quantum Hamming bound from Sec,~\ref{quditrotationerrors} can potentially be violated when protection against photon-loss errors, rather than qudit-rotation errors, accounts for the primary error mechanism.  Our QEC codes are designed to take advantage of this possibility.  

Consider encoding $k$ logical qubits in $n$ physical qutrits when loss of a single photon is the dominant error mechanism.  There are three possible single-photon-loss errors for each qutrit: a single photon may be lost from either the signal, idler, or pump modes. For a single-photon loss from any one of the $3n$ bosonic modes to be correctable, then the total dimension of photon-loss errors,
\begin{align}
D_{\text{err}} =  \sum_{j=0}^1 \binom{3n}{j}2^k,
\end{align}
cannot exceed the dimension, $D_{\text{loss}}$, of the resulting corrupted code subspace.  
Take any qutrit-qubit $\chi^{(2)}$ QEC code as an example.  Its code subspace lies in $\mathcal{H}_2 = {\rm Span}\{\ket{j,j,2-j}: 0\le j \le 2\}$, hence its corrupted code subspace---after loss of a single photon---lies in
\begin{align}
\mathcal{H}_2^\prime=&\text{Span}\{\ket{0,0,2}), \ket{1,1,1},\ket{2,2,0}, \ket{0,0,1},\nonumber \\
&\ket{0,1,1},\ket{1,0,1},\ket{1,1,0},\ket{1,2,0},\ket{2,1,0}\},
\end{align}
implying $D_{\text{loss}}\le 9^n$ and the following photon-loss quantum Hamming bound for qutrit-qubit $\chi^{(2)}$ QEC codes:
\begin{align}
 \sum_{j=0}^1 \binom{3n}{j}2^k  = 2(1+3n) \leq 9^n.
\label{EECCbound}
\end{align}

This bound beats the corresponding generalized quantum Hamming bound, (\ref{GQHB2}), for physical-qudit rotation errors as (\ref{EECCbound}) is satisfied for all $n \ge 1$, whereas (\ref{GQHB2}) requires $n \ge 4$.   
So, because only one physical qutrit is required by the quantum Hamming bound to protect a logical qubit against single-photon-loss errors, we have shown that our qutrit-qubit $\chi^{(2)}$ EECC saturates (\ref{EECCbound}). The following theorem provides the photon-loss generalized quantum Hamming bound on $\chi^{(2)}$ QEC codes that use $n$ physical qudits of dimension $q$ to encode $k$ logical qudits of dimension $b$.\\ 

\noindent\textbf{Theorem 5.} The generalized quantum Hamming bound for single-photon loss errors is $(1+3n)b^k \le (4q-3)^n$.
\\

\noindent\textit{Proof:} Paralleling the derivation of the photon-loss dimension, $D_{\text{err}}$, for encoding of qubits, we have that
\begin{align}
D_{\text{err}} =  \sum_{j=0}^1 \binom{3n}{j}b^k = (1+3n)b^k.
\end{align}
For all of these errors to be correctable by the $\chi^{(2)}$ QEC code, $D_{\text{err}}$ cannot exceed the dimension, $D_{\text{loss}}$, of the corrupted code subspace.  Prior to a photon loss, each physical qudit comes from a code subspace that lies within $\mathcal{H}_{q-1} = {\rm Span}\{\ket{j,j,q-1-j}: 0\le j \le q-1\}$.  For $1 \le j \le q-2$, loss of a single photon from $\ket{j,j,q-1-j}$ corrupts $\mathcal{H}_{q-1}$ by adding three new dimensions., whereas loss of a single photon from $\ket{0,0,q-1}$ corrupts $\mathcal{H}_{q-1}$ by adding one new dimension, and loss of a single photon from $\ket{q-1,q-1,0}$ corrupts $\mathcal{H}_{q-1}$ by adding two new dimensions.  Thus we get $D_{\text{loss}} \le (4q-3)^n$ and our proof is complete.\\

The $\chi^{(2)}$ PCC has $k=1$, $q = b \ge 2$, and $n=2$.  It satisfies the photon-loss quantum Hamming bound, as it must, because we have already shown that it can correct all single-photon loss errors. Moreover, the photon-loss quantum Hamming bound for all $q=b \ge 2$ is violated when $n = 1$, so our $\chi^{(2)}$ PCC saturates this bound.  

Likewise, the $\chi^{(2)}$ EECC, which has $k=1$, $q = 2b-1 \ge 3$, and $n=1$, satisfies the quantum Hamming bound, as it must, because we have shown that it can correct all single-photon loss errors.  Indeed, it saturates this bound.  


\section{Conclusions}
\label{Conclusion}

We have used a stabilizer-inspired symmetry-operator analysis to design three hardware-efficient QEC codes for $\chi^{(2)}$ quantum computation:  the $\chi^{(2)}$ PCC, the $\chi^{(2)}$ EECC, and the $\chi^{(2)}$ BC.  The $\chi^{(2)}$ PCC and $\chi^{(2)}$ EECC need only coherent $\chi^{(2)}$ interactions and linear-optics transformations for their encoding, decoding, and error-correction operations, and their universal encoded-basis gate sets.  Coherent three-wave-mixing in superconducting resonators, together with its nondemolition photon-number parity measurements~\cite{Devoret2010,Schackert2013,Sun2014}, provide what is currently the most promising experimental platform for implementing the $\chi^{(2)}$ PCC and $\chi^{(2)}$ EECC in either the qubit or qutrit bases. Our $\chi^{(2)}$ BC encodes each logical qubit using an average of $3(N-1/2)$ photons and can correct $m$-photon ($m\le N$) loss or gain errors, and $(m-1)$th-order ($m\le N)$ dephasing errors. It is the first known bosonic code with $O(N)$ scaling for the number of photons needed to correct such errors. This scaling advantage comes with a price:  the $\chi^{(2)}$ BC requires channel monitoring that identifies $m$.  Our results thus establish a route to breaking the existing ceiling on encoding efficiency by including new measurement strategies.   We have also derived generalized quantum Hamming bounds for $\chi^{(2)}$ QEC codes and for nondegenerate codes that correct photon-loss errors.  The $\chi^{(2)}$ PCC and the $\chi^{(2)}$ EECC were shown to saturate their respective photon-loss quantum Hamming bounds. Notably, our symmetry-operator framework provides a systematic way for constructing bosonic QEC codes based on properties of the underlying system dynamics. It  also provides  a straightforward generalization from qubit-basis three-mode encoding to qudit-basis multi-mode encoding.  

\section*{Acknowledgements}

M. Y. N. and J. H. S. acknowledge support from Air Force Office of Scientific Research Grant No. FA9550-14-1-0052. M. Y. N. acknowledges  support from Claude E. Shannon Research Assistantship.  I. L. C. acknowledges support from the National Science Foundation Center for Ultracold Atoms. We thank   Theodor Yoder, Liang Jiang, Michel Devoret, Victor V. Albert and Kyungjoo Noh for enlightening discussions.


\begin{appendix}

\section{Encoding, Decoding, and Error Correction for Qutrit-Basis $\chi^{(2)}$ Parity-Check Code}\label{ENCODE}

In Appendix~\ref{ENCODE}, we present the encoding, decoding, error-detection, and error-correction procedures for the qutrit-basis $\chi^{(2)}$ PCC given in Eqs.~(\ref{EC1}-\ref{EC3}); Appendix~\ref{GATE} provides a similar development for the qubit-basis $\chi^{(2)}$ EECC.\\

\noindent{\bf Encoding}. Encoding of the qutrit-basis  $\chi^{(2)}$ PCC amounts to preparation of the code's logical-zero state $\ket{\tilde{0} } =  \ket{111}_1\ket{111}_2$, because any quantum computation can be decomposed into a sequence of universal gates  acting on the all-zero logical state.   The $\chi^{(2)}$ PCC's logical-zero state contains a single photon in each of its six bosonic modes, which can be prepared by combining the heralded single photons---of the appropriate frequencies and polarizations---generated by incoherent (strong, nondepleting pump) SPDC processes into one spatial mode by means of dichroic mirrors and polarizing beam-splitters.\\

\noindent{\bf Decoding}.  Decoding for the qutrit-basis $\chi^{(2)}$ PCC can be realized as follows.  First, we by apply the qutrit-CNOT gate 
\begin{widetext}
\begin{align}
\text{CNOT}^3_{1,2}=&\ket{1,1,1}_{1\,1}\bra{1,1,1}\otimes \hat{I}_2 +  \ket{0,0,2}_{1\,1}\bra{0,0,2}\otimes(\ket{2,2,0}_{2\,2}\bra{1,1,1} + \ket{1,1,1}_{2\,2}\bra{0,0,2} +\ket{0,0,2}_{2\,2}\bra{2,2,0})  \nonumber \\ 
&+ \ket{2,2,0}_{1\,1}\bra{2,2,0}\otimes (\ket{0,0,2}_{2\,2}\bra{1,1,1} + \ket{1,1,1}_{2\,2}\bra{2,2,0} + \ket{2,2,0}_{2\,2}\bra{0,0,2}),
\end{align}
\end{widetext} where the superscript 3 denotes a qutrit-basis gate and the subscripts 1,2 indicate that the first physical-basis state is the control and while the second is the target.
This gate transforms the logical basis states $\{\ket{\tilde{j}}, j = 0,1,2\}$ into $\{\ket{\tilde{j}'},j =0,1,2\}$ given by
\begin{align}
\ket{\tilde{2}^\prime}=&(\ket{2,2,0}_1\ket{2,2,0}_2 + \ket{0,0,2}_1\ket{0,0,2}_2)/\sqrt{2}, \\
\ket{\tilde{1}^\prime}=&(\ket{2,2,0}_1+\ket{0,0,2}_1)\ket{1,1,1}_2/\sqrt{2},\\ 
 \ket{\tilde{0}^\prime }= &\ket{1,1,1}_1\ket{1,1,1}_2.
\end{align}
Next, we apply the CNOT$^3_{2,1}$ gate to transform the $\{\ket{\tilde{j}'}\}$ into $\{\ket{\tilde{j}''}\}$ given by
\begin{align}
\ket{\tilde{2}^{\prime\prime}}=&\ket{1,1,1}_1(\ket{2,2,0}_2  +\ket{0,0,2}_2)/\sqrt{2},\\
\ket{\tilde{1}^{\prime\prime}}=&(\ket{2,2,0}_1  +\ket{0,0,2}_1)\ket{1,1,1}_2/\sqrt{2},\\ 
\ket{\tilde{0}^{\prime\prime}}=&\ket{1,1,1}_1\ket{1,1,1}_2.
\end{align}
Computational-basis measurements can now be completed by making photon-number resolving measurements on the six bosonic modes:  $\ket{\tilde{0}}$ is identified by every mode containing a single photon; $\ket{\tilde{1}}$ is identified by only the second physical qutrit having modes containing single photons; and $\ket{\tilde{2}}$ is identified by only the first physical qutrit having modes  containing single photons. \\ 

\noindent{\bf Error Detection}.  Error detection for the $\chi^{(2)}$ PCC uses nondemolition measurements of the $\chi^{(2)}$ PCC's photon-number parity vector, ${\bf p}_{12}$ from Eq.~(\ref{ParityVector}), and its generalized photon-number parity vector, ${\bf q}_{12}$ from Eq.~(\ref{q12eq}), to obtain the unique syndromes
\begin{align}
\hat{a}_{s_1} &\to {\bf p}_{12}=(1,1,0,0,0,0), \, {\bf q}_{12}=(2,0),\\
\hat{a}_{i_1} &\to {\bf p}_{12}=(1,0,1,0,0,0) ,  \, {\bf q}_{12}=(2,0),\\
\hat{a}_{p_1} &\to {\bf p}_{12}=(0,1,1,0,0,0), \, {\bf q}_{12}=(2,0),\\
\hat{a}_{s_2} &\to {\bf p}_{12}=(0,0,0,1,1,0),  \, {\bf q}_{12}=( 0,2),\\
\hat{a}_{i_2} &\to {\bf p}_{12}=(0,0,0,1,0,1) ,  \, {\bf q}_{12}=( 0,2),\\
\hat{a}_{p_2} &\to {\bf p}_{12}=(0,0,0,0,1,1),  \, {\bf q}_{12}=( 0,2),
\end{align}
for photon-loss errors, and
\begin{align}
\hat{a}_{s_1}^\dagger &\to {\bf p}_{12}=(1,1,0,0,0,0),\,  {\bf q}_{12}=(1,0),\\
\hat{a}_{i_1}^\dagger &\to {\bf p}_{12}=(1,0,1,0,0,0),\,  {\bf q}_{12}=(1,0),\\
\hat{a}_{p_1}^\dagger &\to {\bf p}_{12}=(0,1,1,0,0,0),\,  {\bf q}_{12}=(1,0),\\
\hat{a}_{s_2}^\dagger &\to {\bf p}_{12}=(0,0,0,1,1,0),\,  {\bf q}_{12}=(0,1),\\
\hat{a}_{i_2}^\dagger &\to {\bf p}_{12}=(0,0,0,1,0,1),\,  {\bf q}_{12}=(0,1),\\
\hat{a}_{p_2}^\dagger &\to {\bf p}_{12}=(0,0,0,0,1,1),\,  {\bf q}_{12}=(0,1).
\end{align}
for photon-gain errors.\\

\noindent{\bf Error Correction}.  Single-photon loss errors can be corrected using just linear optics and $\chi^{(2)}$ computational primitives, as we now explain.  To correct a single-photon-loss error, we  must increase the photon number of the corrupted  mode  by one, while to correct  a single-photon gain error, we must decrease the photon number of the corrupted mode by one.   Because single-photon gain errors are much less likely to occur than single-photon loss errors in $\chi^{(2)}$ media and linear-optical circuits, we will only provide a specific error-correction circuit for single-photon loss errors. Without loss of generality, we will assume that the single-photon loss error is in one of the first physical qutrit's modes.    

Our error-correction procedure presumes that the $\chi^{(2)}$ PCC's first and second qutrits are single-rail encoded on different rails.  For both qutrits, the signal and idler photons are frequency degenerate ($\omega_s = \omega_i = \omega$) and orthogonally polarized, while the pump photons have frequency $\omega_p = 2\omega$ and are co-polarized with those of the idler.  Consequently, signal, idler, and pump photons that are propagating on a single rail can be directed to separate rails---using a polarizing beam splitter~(PBS) and a dichroic mirror~(DM)---for individual processing, after which they can be recombined on a single rail using those same linear-optics resources.  We now present the error-correction steps for single-photon loss error on the first qutrit's signal mode (Case A) and the first qutrit's pump mode (Case B).\\ 
 
\begin{figure}[h]
\begin{center}
\includegraphics[width=1\linewidth]{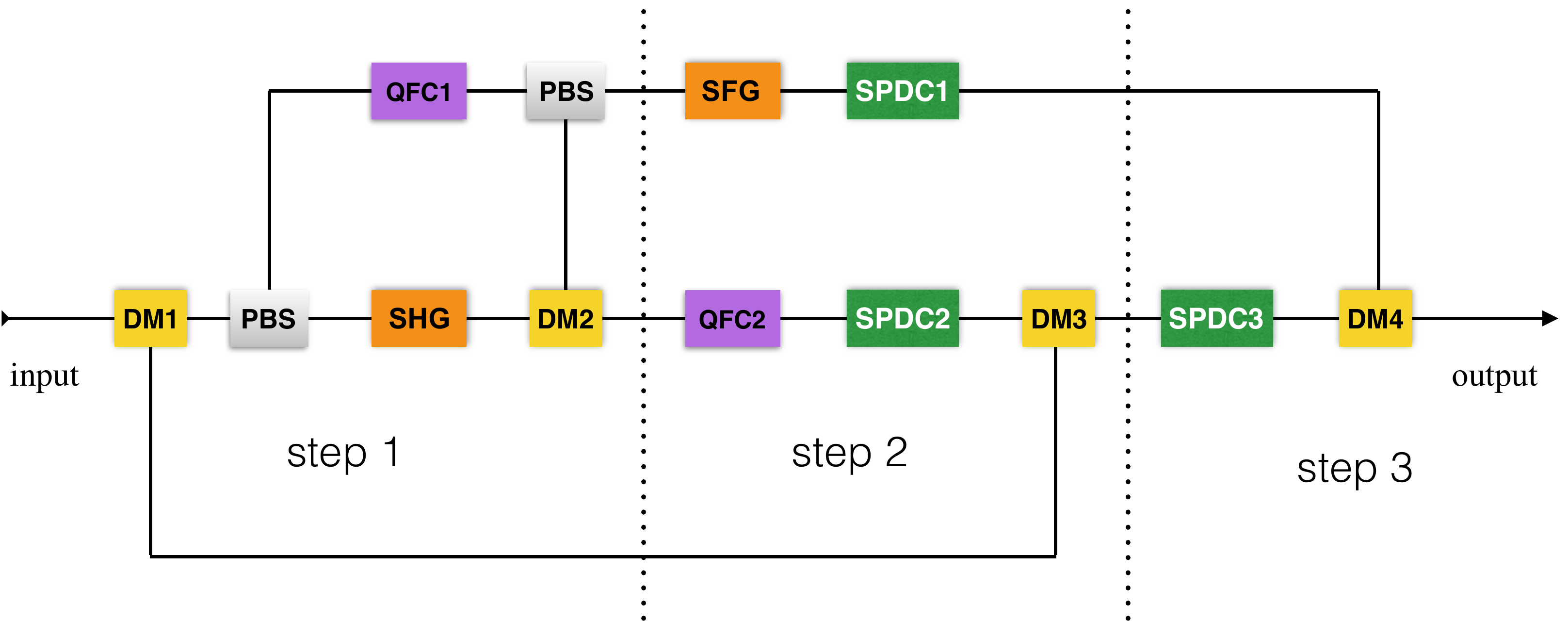} 
\caption{Circuit for restoring the two-pump-photon subspace after loss of a single signal-mode photon from the $\chi^{(2)}$ PCC's first qutrit.  Input is the encoded state's first qutrit, which has suffered a signal-photon loss; output is the encoded state's first qutrit restored to the two-pump-photon subspace.  DM1, DM2, DM3, DM4:  dichroic mirrors. PBS:  polarizing beam-splitter.  SHG: second-harmonic generation.  QFC1, QFC2:  quantum-state frequency conversion.  SFG:  sum-frequency generation.  SPDC1, SPDC2, SPDC3:  frequency-degenerate, type-I phase-matched spontaneous parametric downconversion. See text for details.\label{EC}}
\end{center}
\end{figure}

\noindent\underline{Case A: Correction of a First-Qutrit Signal-Photon Loss}.  

Loss of a signal photon from the $\chi^{(2)}$ PCC's first qutrit converts the encoded qutrit  $\ket{\psi_0}=\alpha\ket{\tilde{2}}  +\beta \ket{\tilde{1} }  +\gamma\ket{\tilde{0}}$ to 
\begin{align}
\ket{\psi_0^\prime} = &\alpha\ket{1,2,0}_1\ket{0,0,2}_2 +\beta \ket{1,2,0}_1\ket{2,2,0}_2 \nonumber \\
& +\gamma\ket{0,1,1}_1\ket{1,1,1}_2,  
\label{psi0prime}
\end{align}
which shows that first physical basis is no longer in the two-pump-photon subspace.  Error correction is accomplished by transforming $\ket{1,2,0}_1$ to  $\ket{2,2,0}_1$ and $\ket{0,1,1}_1$ to $\ket{0,0,2}_1$, to restore the qutrit to the two-pump-photon subspace, and then applying a sequence of gates realized with $\chi^{(2)}$ computational primitives, to restore the encoded basis back to that defined in Eqs.~(\ref{EC1})--(\ref{EC3}). We restore the physical basis to the two-pump-photon subspace using the optical circuit shown in Fig.~\ref{EC}, whose three steps are described below. \\ 

\noindent\textit{Step~1.} Dichroic mirror DM1 directs the first qutrit's pump photons from the input rail to the bottom rail in Fig.~\ref{EC}. Polarizing  beam-splitter PBS then directs signal photons from the original rail to the top rail in Fig.~\ref{EC}.  There, a quantum-state frequency conversion (QFC1) converts a frequency-$\omega$ single-photon Fock-state signal, if present, to a frequency-$2\omega$ single-photon Fock state.  Meanwhile, a second-harmonic generation (SHG1) in the middle rail converts a two-photon Fock-state idler, if present, to a frequency-$2\omega$ single-photon Fock state that is co-polarized with the signal.  Until this point, the $\ket{0,1,1}_1$ component of $\ket{\psi_0'}$ has been unaffected, but its $\ket{1,2,0}_1$ component has been transformed to $\ket{0,0,1}_1\ket{1}_{s}$, where $\ket{1}_{s}$ to denote a frequency-$2\omega$ single-photon Fock state that is co-polarized with the signal.  Dichroic mirror DM2 then directs the frequency-$2\omega$ photon from the middle rail, if present, to the upper rail through a PBS.  Thus, at the end of Step~1, the $\ket{0,0,1}_1\ket{1}_{s}$ state component is confined to the top rail, while the idler photon of the $\ket{0,1,1}_1$ state component resides in the middle rail, and that state component's pump photon occupies the bottom rail.\\ 
 
\noindent\textit{Step~2.} In Step~2, a sum-frequency generation (SFG) stage in the top rail first converts  $\ket{0,0,1}_1\ket{1}_{s}$ into a frequency-$4\omega$ single-photon Fock state that is co-polarized with the signal, after which the frequency-degenerate, type-I phase-matched SPDC1 transforms that single-photon Fock state to $\ket{0,0,2}_1$. In Step~2's middle rail, the quantum-state frequency conversion QFC2 coherently converts the frequency-$\omega$ idler photon to frequency $2\omega$, from which the frequency-degenerate, type-I phase-matched SPDC2 produces a two-photon Fock state in the idler mode.  SPDC2's output is then combined with the bottom rail's pump photon via dichroic mirror DM3, thus completing conversion of $\ket{0,1,1}_1$---which had been distributed between the middle and bottom rails---to $\ket{0,2,1}_1$ residing on the middle rail.\\ 

\noindent\textit{Step~3.}  Step~3 completes restoration of the two-pump-photon subspace as follows.  It first uses the frequency-degenerate, type-I phase-matched SPDC3 in the middle rail to convert a single pump-mode photon into a pair of signal-mode photons, realizing the transformation of $\ket{0,2,1}_1$ to $\ket{2,2,0}_1$. Then it employs dichroic mirror DM4 to recombine the top-rail's $\ket{0,0,2}_1$ contribution with the middle rail's $\ket{2,2,0}_1$ contribution so that when $|\psi'_0\rangle$ from Eq.~(\ref{psi0prime}) is the input to the Fig.~\ref{EC} circuit, the output state that results is
\begin{align}
\ket{\psi_1}=  &\alpha\ket{0,0,2}_1\ket{0,0,2}_2
+\beta \ket{0,0,2}_1\ket{2,2,0}_2 \nonumber \\ 
 &+\gamma\ket{2,2,0}_1\ket{1,1,1}_2.
\end{align}

Having restored the $\chi^{(2)}$ PCC's first qutrit to the two-pump-photon subspace, correcting for that qutrit's loss of a single signal-mode photon is completed by applying the following gates.  \\

\noindent\textit{Gate~1.} The first gate we apply is the qubit CNOT gate,
\begin{align}
&\text{CNOT}^2_{2,1} = \hat{I}_1\otimes(\ket{0,0,2}_{2\,2}\bra{0,0,2} + \ket{2,2,0}_{2\,2}\bra{2,2,0}) \nonumber \\
&\quad+(\ket{0,0,2}_{1\,1}\bra{0,0,2}+\ket{1,1,1}_{1\,1}\bra{2,2,0} \nonumber \\
&\quad + \ket{2,2,0}_{1\,1}\bra{1,1,1}) \otimes \ket{1,1,1}_{2\,2}\bra{1,1,1}/\sqrt{2},
\label{CNOT221}
\end{align}
which gives
\begin{align}
&\ket{\psi_2}= \text{CNOT}_{2,1}^2 \ket{\psi_1}= \alpha\ket{0,0,2}_1\ket{0,0,2}_2\nonumber\\ 
& \quad +\beta \ket{0,0,2}_1\ket{2,2,0}_2+\gamma\ket{1,1,1}_1\ket{1,1,1}_2.
\end{align}

\noindent\textit{Gate~2.} The second gate we need is the controlled-Hadamard gate,
\begin{align}
&\Lambda_{21}(H) = \hat{I}_1\otimes(\ket{0,0,2}_{2\,2}\bra{0,0,2}+\ket{1,1,1}_{1\,1}\bra{1,1,1})
\nonumber \\
&+(\ket{+}_{1\,1}\bra{0,0,2}+\ket{-}_{1\,1}\bra{2,2,0})\otimes\ket{2,2,0}_{2\,2}\bra{2,2,0},
\label{lam21}
\end{align}
where $\ket{\pm}_1 \equiv (\ket{0,0,2}_1\pm \ket{2,2,0}_1)/\sqrt{2}$, which gives 
\begin{align}
\ket{\psi_3}&=\Lambda_{21}(H) \ket{\psi_2} =\alpha \ket{0,0,2}_1\ket{0,0,2}_2 \nonumber\\ 
&\quad+\beta\ket{+}_1\ket{2,2,0}_2 +\gamma\ket{1,1,1}_1\ket{1,1,1}_2.
\end{align} 

\noindent\textit{Gate~3.} The third gate we use is the not-controlled  Hadamard gate, 
\begin{align}
&\bar{\Lambda}_{21}(H) = \hat{I}_1\otimes(\ket{1,1,1}_{1\,1}\bra{1,1,1}+\ket{2,2,0}_{2\,2}\bra{2,2,0})
\nonumber \\
&+(\ket{+}_{1\,1}\bra{0,0,2}+\ket{-}_{1\,1}\bra{2,2,0})\otimes\ket{0,0,2}_{2\,2}\bra{0,0,2},
\label{notlam21}
\end{align}
which gives
\begin{align}
\ket{\psi_4}&=\bar{\Lambda}_{21}(H) \ket{\psi_3} =\alpha\ket{+}_1\ket{0,0,2}_2 \nonumber \\ 
&\quad+\beta\ket{+}_1\ket{2,2,0}_2 +\gamma\ket{1,1,1}_1\ket{1,1,1}_2.
\end{align}

\noindent\textit{Gate~4.}  The final gate we need is the qubit-CNOT gate
\begin{align}
\text{CNOT}^{2\prime}_{1,2} &= (\ket{1,1,1}_{1\,1}\bra{1,1,1}+\ket{2,2,0}_{1\,1}\bra{2,2,0})\otimes \hat{I}_2
\nonumber\\
&\quad + \ket{0,0,2}_{1\,1}\bra{0,0,2}\otimes(\ket{2,2,0}_{2\,2}\bra{0,0,2} \nonumber\\
&\quad + \ket{0,0,2}_{2\,2}\bra{2,2,0}).
\label{CNOTprime}
\end{align}
We then get
\begin{align}
\ket{\psi_5} &= \text{CNOT}^{2\prime}_{1,2}\ket{\psi_4} \\\nonumber
&=\alpha(\ket{2,2,0}_1 \ket{0,0,2}_2+ \ket{0,0,2}_1 \ket{2,2,0}_2)/\sqrt{2}  \\\nonumber
&+\beta(\ket{2,2,0}_1\ket{2,2,0}_2+ \ket{0,0,2}_1\ket{0,0,2}_2)/\sqrt{2} \nonumber \\
&+\gamma\ket{1,1,1}_1\ket{1,1,1}_2,
\end{align}
which completes recovery from loss of a first-qutrit signal photon, because $\ket{\psi_5} = \ket{\psi_0}$.\\

\noindent\underline{Case B: Correction of a First-Qutrit Pump-Photon Loss}.  We correct loss of a pump photon in the $\chi^{(2)}$ PCC's first qutrit by a procedure similar to that in Case~A, i.e., we first coherently increase the photon number in the corrupted pump mode by one, to restore the physical basis to the two-pump-photon subspace, and then apply a sequence of gates realized with $\chi^{(2)}$ computational primitives, to restore the encoded basis back to that defined in Eqs.~(\ref{EC1})--(\ref{EC3}).

\begin{figure}[h]
\begin{center}
\includegraphics[width=1\linewidth]{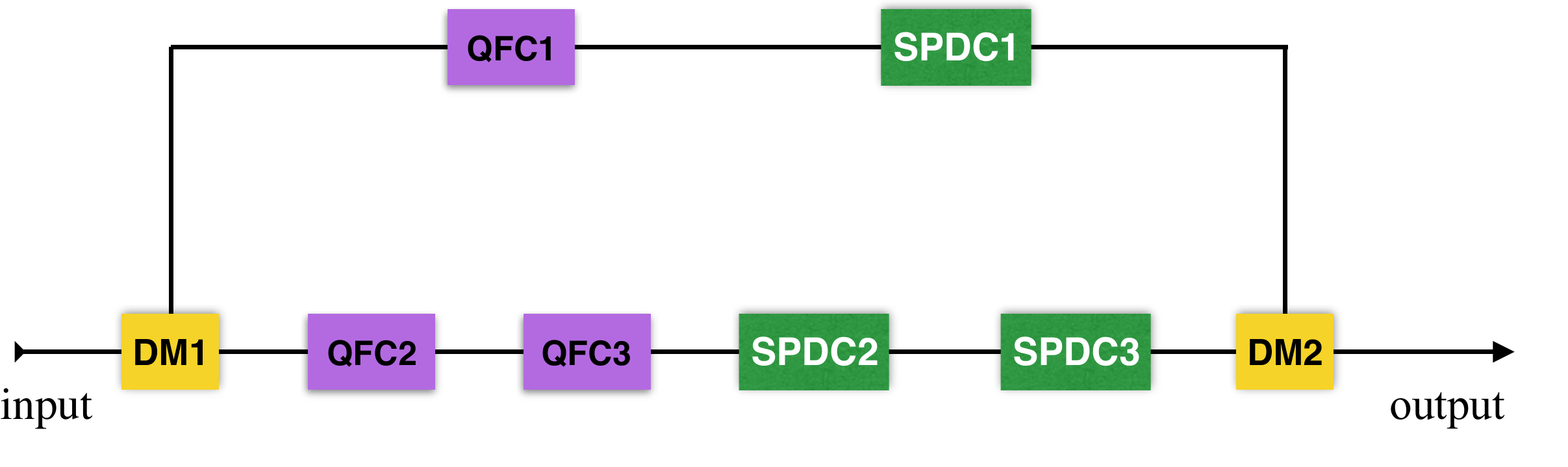}
\caption{Circuit for restoring the two-pump-photon subspace after loss of a single pump-mode photon from the $\chi^{(2)}$ PCC's first qutrit.  Input is the encoded state's first qutrit, which has suffered a pump-photon loss; output is the encoded state's first qutrit restored to the two-pump-photon subspace. DM1, DM2:  dichroic mirrors.  QFC1, QFC2, QFC3:  quantum-state frequency conversion.  SPDC1, SPDC2:  frequency-degenerate, type-0 phase-matched spontaneous parametric downconversion. SPDC3 :  frequency-degenerate, type-I phase-matched spontaneous parametric downconversion. See text for details. \label{ECpump}}
\end{center}
\end{figure}

Loss of a first-qutrit pump photon from the encoded qutrit $\ket{\psi_0}=\alpha\ket{\tilde{2}}  +\beta \ket{\tilde{1} }  +\gamma\ket{\tilde{0}}$ results in 
\begin{align}
\ket{\psi_0^\prime}=&\alpha\ket{0,0,1}_1\ket{2,2,0}_2 +\beta \ket{0,0,1}_1\ket{0,0,2}_2 \nonumber \\ 
&+\gamma \ket{1,1,0}_1\ket{1,1,1}_2.
\label{pumplossstate}
\end{align}
The optical circuit shown in Fig.~\ref{ECpump}, which restores the physical basis to the two-pump-photon subspace, works as follows.  Dichroic mirror DM1 directs the pump photon to the top rail in Fig.~\ref{ECpump} leaving the signal and idler photons on the original rail, after which the $\ket{0,0,1}_1$ component of the first qutrit resides on the top rail, while the $\ket{1,1,0}_1$ component remains on the original rail.  Quantum-state frequency conversion QFC1 coherently converts the frequency-$2\omega$ pump photon on the top rail to frequency $4\omega$ from which the frequency-degenerate, type-0 phase-matched SPDC1 produces a two-photon Fock state in the pump mode. At this point, the top rail contains the $\ket{0,0,2}_1$ component of the overall state.  

On the original rail, quantum-state frequency conversions QFC2 and QFC3 are phase matched so that QFC2 coherently converts a frequency-$\omega$ signal photon to frequency $2\omega$ and QFC3 coherently converts a frequency-$\omega$ idler photon to frequency $2\omega$.  The frequency-degenerate, type-0 phase-matched SPDC2 transforms the frequency-$2\omega$ signal-polarization photon into a signal-mode two-photon Fock state.  Likewise, the frequency-degenerate, type-I phase-matched SPDC3 transforms the frequency-$2\omega$ idler-polarization photon into an idler-mode two-photon Fock state.  After these transformations the original rail contains the $\ket{2,2,0}_1$ component of the overall state.  

After dichroic mirror DM2 combines the outputs of SPDC1 and SPDC3 on the original rail, 
we see that when $\ket{\psi_0'}$ from Eq.~(\ref{pumplossstate}) is the input to the Fig.~\ref{ECpump} circuit, the resulting output state is
\begin{align}
\ket{\psi_1} = & \alpha\ket{0,0,2}_1\ket{2,2,0}_2 \\\nonumber
&+\beta \ket{0,0,2}_1\ket{0,0,2}_2 +\gamma \ket{2,2,0}_1\ket{1,1,1}_2,
\end{align}
which lies in the two-pump-photon subspace.

Error correction for loss of a first qutrit's pump photon is completed by applying the following four-gate sequence.\\

\noindent\textit{Gate~1.}  The first gate we apply is the controlled Hadamard from Eq.~(\ref{lam21}), which gives
\begin{align}
\ket{\psi_2}&=\Lambda_{21}(H) \ket{\psi_1} 
=\alpha \ket{+}_1\ket{2,2,0}_2 \nonumber \\
&\quad+\beta\ket{0,0,2}_1\ket{0,0,2}_2 +\gamma\ket{2,2,0}_1\ket{1,1,1}_2.
\end{align} 

\noindent\textit{Gate~2.} The second gate we use is the not-controlled  Hadamard from Eq.~(\ref{notlam21}), which gives
\begin{align}
\ket{\psi_3}&=\bar{\Lambda}_{21}(H) \ket{\psi_2} =\alpha\ket{+}_1\ket{2,2,0}_2 \nonumber \\
&\quad+\beta\ket{+}_1\ket{0,0,2}_2 +\gamma\ket{2,2,0}_1\ket{1,1,1}_2.
\end{align}

\noindent\textit{Gate~3.} The third gate we employ is CNOT$^2_{2,1}$ from Eq.~(\ref{CNOT221}), which gives
\begin{align}
\ket{\psi_4}&=\text{CNOT}^2_{2,1}\ket{\psi_3} =\alpha\ket{+}_1\ket{2,2,0}_2 \nonumber \\
&\quad+\beta\ket{+}_1\ket{0,0,2}_2 +\gamma\ket{1,1,1}_1\ket{1,1,1}_2.
\end{align}

\noindent\textit{Gate~4.} The last gate we  employ is 
\begin{align}
\text{CNOT}^{2\prime\prime}_{1,2} &= (\ket{0,0,2}_{1\,1}\bra{0,0,2}+\ket{1,1,1}_{1\,1}\bra{1,1,1})\otimes \hat{I}_2
\nonumber\\
&\quad + \ket{2,2,0}_{1\,1}\bra{2,2,0}\otimes(\ket{2,2,0}_{2\,2}\bra{0,0,2} \nonumber\\
&\quad + \ket{0,0,2}_{2\,2}\bra{2,2,0}).
\label{CNOTprimeprime}
\end{align}
which gives
\begin{align}
\ket{\psi_5} &= \text{CNOT}^{2\prime\prime}_{1,2}\ket{\psi_4} \\\nonumber
&=\alpha(\ket{2,2,0}_1 \ket{0,0,2}_2+ \ket{0,0,2}_1 \ket{2,2,0}_2)/\sqrt{2}  \\\nonumber
&+\beta(\ket{2,2,0}_1\ket{2,2,0}_2+ \ket{0,0,2}_1\ket{0,0,2}_2)/\sqrt{2} \nonumber \\
&+\gamma\ket{1,1,1}_1\ket{1,1,1}_2,
\end{align}
and completes the  recovery from loss of a first-qutrit pump photon.

\section{Encoding, Decoding, and Error Correction for the Qubit-Basis $\chi^{(2)}$  Embedded Error-Correcting Code}
\label{GATE}

Below, we present encoding, decoding. and error correction procedures for the qubit-basis $\chi^{(2)}$  EECC. We also give an explicit procedure---using coherent $\chi^{(2)}$ interactions---for constructing the logical rotation ($\hat{X}_P$) and Hadamard ($\hat{H}$) gates for this encoded qubit, as these gates are used in the error-correction procedures.\\

\noindent{\bf Encoding}.  To prepare the qubit-basis $\chi^{(2)}$ EECC's logical-zero state, $\ket{1,1,1}$, we follow the procedure given earlier for the qutrit-basis $\chi^{(2)}$ PCC's encoding of one of its logical-zero state's qutrits.\\

\noindent{\bf Decoding}. Decoding of the qubit-basis $\chi^{(2)}$ EECC  is realized by  making a photon-number parity measurement on any one of the encoded state's three bosonic modes: the photon-number parity every mode in $\ket{\tilde{0}}$ is even, whereas the photon-number parity of every mode in $\ket{\tilde{1}}$ is odd.\\

\noindent {\bf Error Detection}.
Error detection for the qubit-basis $\chi^{(2)}$ EECC employs nondemolition measurements~\cite{Sun2014}  of the photon-number parity vector, ${\bf p}$ from Eq.~(\ref{pvectdefn}), and the generalized photon-number parity, $q_{\rm EECC}$ from Eq.~(\ref{genparvec}).  These measurements provide the following unique syndromes,
\begin{align}
\hat{a}_{s} \to {\bf p}=(1,1,0),\, q_{\rm EECC} =2,\\
\hat{a}_{i} \to {\bf p}=(1,0,1) ,\, q_{\rm EECC} =2,\\
\hat{a}_{p} \to {\bf p}=(0,1,1),\,q_{\rm EECC} =2,
\end{align}
for photon-loss errors, and
\begin{align}
\hat{a}_{s}^\dagger \to {\bf p}=(1,1,0),\, q_{\rm EECC} =1,\\
\hat{a}_{i}^\dagger \to {\bf p}=(1,0,1),\, q_{\rm EECC}=1,\\
\hat{a}_{p}^\dagger \to {\bf p}=(0,1,1),\, q_{\rm EECC} =1,
\end{align}
for photon-gain errors.\\ 

\noindent {\bf Logical rotation and Hadamard gates}.  Here we show that the $\hat{X}_P$ and $\hat{H}$ gates, defined in Eqs.~(\ref{XP}) and (\ref{Had}), are realizable with unitary transformations generated by $\chi^{(2)}$ Hamiltonians. For that purpose, we augment the $\chi^{(2)}$ Hamiltonians $\hat{G}_1,\hat{G}_2$, from Eqs.~(\ref{HSPDC11}) and (\ref{HSPDC}), with five additional $\chi^{(2)}$ generators, $\{\hat{G}_k :3\le k\le 7\}$, from  the $\mathfrak{u}(3)$ Lie algebra in the two-pump-photon subspace $\mathcal{H}_2=\text{Span}\{\ket{0,0,2},\ket{1,1,1},\ket{2,2,0}\}$~\cite{Niu2017} that are obtained as follows:
\begin{align}\label{G3}
&\hat{G}_3=i[\hat{G}_1, \hat{G}_2]=\begin{bmatrix}
		1 &0&0\\
	0  & -2&0\\
	0& 0&1
\end{bmatrix},\\\label{Goriginal4}
&\hat{G}_4= i[\hat{G}_2,\hat{G}_3]=3\begin{bmatrix}
		0 &1&0\\
	1  & 0&0\\
	0&0&0
\end{bmatrix},\\
& \hat{G}_5= i[\hat{G}_3,\hat{G}_1]=3i\begin{bmatrix}
		0 &1&0\\
	-1  & 0&0\\
	0&0&0
\end{bmatrix},\\
&\hat{G}_6=\frac{1}{2}\left( i[\hat{G}_1, \hat{G}_4] +i[\hat{G}_5, \hat{G}_2] \right)=\frac{3}{4}\begin{bmatrix}
		0 &0&0\\
	0  & 0&1\\
	0&1&0
\end{bmatrix},\\\label{G7}
& \hat{G}_7=i[\hat{G}_4, \hat{G}_2]=\frac{3i}{4}\begin{bmatrix}
		0 &0&0\\
	0  & 0&-1\\
	0&1&0
\end{bmatrix}.
\end{align}
Here, we have set $\kappa = 1$ in Eqs.~(\ref{HSPDC11}) and (\ref{HSPDC}), and the matrix representations, which only apply in the two-pump-photon subspace $\mathcal{H}_2$, employ the ${\bf v}^T= [\begin{array}{ccc} v_1 & v_2, & v_3 \end{array}]$ basis for $\mathcal{H}_2$ in which an arbitrary pure-state qutrit is $\ket{\psi}=v_1\ket{1,1,1}+ v_2 \ket{2,2,0} +v_3\ket{0,0,2}$. 

Using the preceding generators, the $\hat{X}_P$ gate can be implemented by the unitary evolution up to a global phase
\begin{align}\label{XPGate}
\hat{X}_P=e^{i2\pi\hat{G}_6/3} e^{i\pi\hat{G}_7/3},
\end{align}
and the $\hat{H}'$ gate can be realized via
\begin{align}
\hat{H}^\prime=e^{i\pi\hat{G}_4/6} e^{-i\pi\hat{G}_5/12}
\end{align}
up to a global phase. Together with Eq.~(\ref{XPGate}) and $\hat{H}= \hat{X}_P^{-1}\hat{H}^\prime\hat{X}_P$, our $\hat{H}'$ result now gives the decomposition of Hadamard gate in the encoded basis as
\begin{align}
\hat{H}=e^{-i\pi\hat{G}_7/3}e^{-i2\pi\hat{G}_6/3} e^{i\pi\hat{G}_4/6} e^{-i\pi\hat{G}_5/12}e^{i2\pi\hat{G}_6/3} e^{i\pi\hat{G}_7/3}.
\end{align}
 
\noindent {\bf Error Correction}.
Similar to what we showed for the qutrit-basis $\chi^{(2)}$ PCC, the qubit-basis $\chi^{(2)}$ EECC's error-correction procedure  can be categorized into two cases:   single-photon loss in either the signal or idler mode, and single-photon loss in the pump mode.   Because of the symmetry between the signal and idler modes, we shall only exhibit error-correction for signal-mode and pump-mode photon losses.  \\

\noindent\underline{Case A:  Correction of a Signal-Photon Loss}.  After a single-photon loss in its signal mode, the pure state $\ket{\psi_0}=\alpha\ket{\tilde{0}} +\beta\ket{\tilde{1}}$ in the qubit-basis $\chi^{(2)}$ EECC becomes
\begin{align}
\ket{\psi_0'}=\alpha\ket{1,2,0} +\beta\ket{0,1,1}.
\end{align}
To restore the original encoded state, we first bring the corrupted state back to the two-pump-photon subspace using the optical circuit shown   in Fig.~\ref{EC}, which yields
\begin{align}
\ket{\psi_1}=\alpha\ket{0,0,2} +\beta\ket{2,2,0}.
\end{align}
Then we apply a unitary gate generated by $\hat{G}_4$ to obtain the state
\begin{align}\label{gate1}
\ket{\psi_2}=e^{i\pi\hat{G}_5/6}\ket{\psi_1}=\alpha\ket{0,0,2} +\beta\ket{1,1,1}.
\end{align}
Finally, we employ a unitary gate generated by $\hat{G}_7$ to recover the original state, 
\begin{align}\label{gate2}
\ket{\psi_3}&=e^{i \pi\hat{G}_7/3}\ket{\psi_2}=\alpha(\ket{0,0,2}+\ket{2,2,0})/\sqrt{2} \nonumber \\
&\quad +\beta\ket{1,1,1} = \ket{\psi_0}.
\end{align}

\noindent\underline{Case B:  Correction of a Pump-Photon Loss}.  After a single-photon loss in its pump mode, the pure state $\ket{\psi_0}=\alpha\ket{\tilde{0}} +\beta\ket{\tilde{1}}$ in the qubit-basis $\chi^{(2)}$ EECC becomes
\begin{align}
\ket{\psi_0'}=\alpha\ket{0,0,1} +\beta\ket{1,1,0}.
\end{align}
To restore the original encoded state, we first bring the corrupted state back to the two-pump-photon subspace using the optical circuit  shown in Fig.~\ref{ECpump}, which produces
\begin{align}
\ket{\psi_1}=\alpha\ket{0,0,2} +\beta\ket{2,2,0}.
\end{align}
Finally, applying the gate sequence defined in Eqs.~(\ref{gate1}) and~(\ref{gate2})  restores the original qubit-basis encoded state.

\end{appendix}


\begin{thebibliography}{99}
\bibitem{footnote}This circuit uses single-rail encoding for the control and target qutrits, i.e., their signal, idler, and pump photons have a common spatial mode, with the signal and idler's photons being orthogonally polarized and the pump's photons being co-polarized with the idler's.
\end{thebibliography}
\end{document}